\documentclass[a4paper, onecolumn]{article}

\usepackage{geometry}
\usepackage[round]{natbib}   % omit 'round' option if you prefer square brackets
\usepackage{authblk}
\usepackage{graphicx}
\usepackage[utf8]{inputenc}
\usepackage[labelfont=bf, figurename=Fig., font={small}]{caption}
\usepackage{cite}
\usepackage[group-four-digits=true,
            output-decimal-marker={.},
            ]{siunitx}
\usepackage{listings}

\usepackage{float}
\usepackage{fancyhdr}
\usepackage{longtable}
\usepackage{multirow}
\usepackage{array}
\newcolumntype{P}[1]{>{\centering\arraybackslash}p{#1}}
\usepackage{xr-hyper}
\usepackage{hyperref}
\hypersetup{
    colorlinks=True,
    linkcolor=blue,
    filecolor=black,      
    urlcolor=blue,
    anchorcolor=black,
    citecolor=black
    }

\usepackage[group-four-digits=true,
            output-decimal-marker={.},
            ]{siunitx}
\usepackage{float}
            
\usepackage{caption}
\usepackage{subcaption}

\date{}

\title{How Good Is Open Bicycle Infrastructure Data? \\A Countrywide Case Study of Denmark}

\author[1]{Ane Rahbek Vierø}
\author[1]{Anastassia Vybornova}
\author[1,2,3]{Michael Szell}
\affil[1]{NEtwoRks, Data, and Society (NERDS), Computer Science Department, IT University of Copenhagen, 2300 Copenhagen, Denmark} 
\affil[2]{ISI Foundation, 10126 Turin, Italy} 
\affil[3]{Complexity Science Hub Vienna, 1080 Vienna, Austria}

\date{}

\begin{document}

\maketitle

\noindent Cycling is a key ingredient for a sustainability shift of Denmark's transportation system. To increase cycling rates, a better nationwide network of bicycle infrastructure is required. Planning such a network requires high-quality infrastructure data, however, the quality of bicycle infrastructure data is severely understudied. Here, we compare Denmark's two largest open data sets on dedicated bicycle infrastructure, OpenStreetMap (OSM) and GeoDanmark, in a countrywide data quality assessment, asking whether data is good enough for network-based analysis of cycling conditions. We find that neither of the data sets is of sufficient quality, and that data set conflation is necessary to obtain a complete dataset. Our analysis of the spatial variation of data quality suggests that rural areas are more likely to suffer from problems with data completeness. We demonstrate that the prevalent method of using infrastructure density as a proxy for data completeness is not suitable for bicycle infrastructure data, and that matching of corresponding features thus is necessary to assess data completeness. Based on our data quality assessment we recommend strategic mapping efforts towards data completeness, consistent standards to support comparability between different data sources, and increased focus on data topology to ensure high-quality bicycle network data.\\


\section{Introduction}

Our current car-dominated transport systems must become more environmentally and socially sustainable \citep{mattioli_chapter_2021, jaramillo_transport_2022, eea_greenhouse_2022}. Active mobility, such as cycling, is an important part of the transition \citep{ec_new_2021, jaramillo_transport_2022, european_commission_european_2023}. However, getting more people to cycle is a complex task and often requires bicycle infrastructure improvements \citep{schoner_missing_2014,buehler_bikeway_2016,xiao_shifting_2022, fosgerau_bikeability_2023}. The corresponding policy and decision-making process could be greatly supported by data-driven methods, as demonstrated by recent bicycle planning approaches \citep{chips_european_2019, eudaly_portland_2020, ecf_integrated_2022} and by active mobility research \citep{lovelace_propensity_2017, natera_orozco_data-driven_2020, olmos_data_2020, steinacker_demand-driven_2022, szell_growing_2022, vybornova_automated_2022, paulsen_societally_2023}. Unfortunately, the often low or unknown quality of bicycle infrastructure data is a massive obstacle for data-driven methods. In terms of quality, cycling data is still lagging behind motorized transport data \citep{lee_emerging_2020, willberg_comparing_2021, ramboll_walking_2022}. Despite this, there is to our knowledge currently no data quality assessment methods tailored specifically to bicycle infrastructure data. All of these issues pose a barrier for any data-informed efforts to improve cycling conditions.

\begin{figure}[H]
\centering
\includegraphics[width=0.999\textwidth]{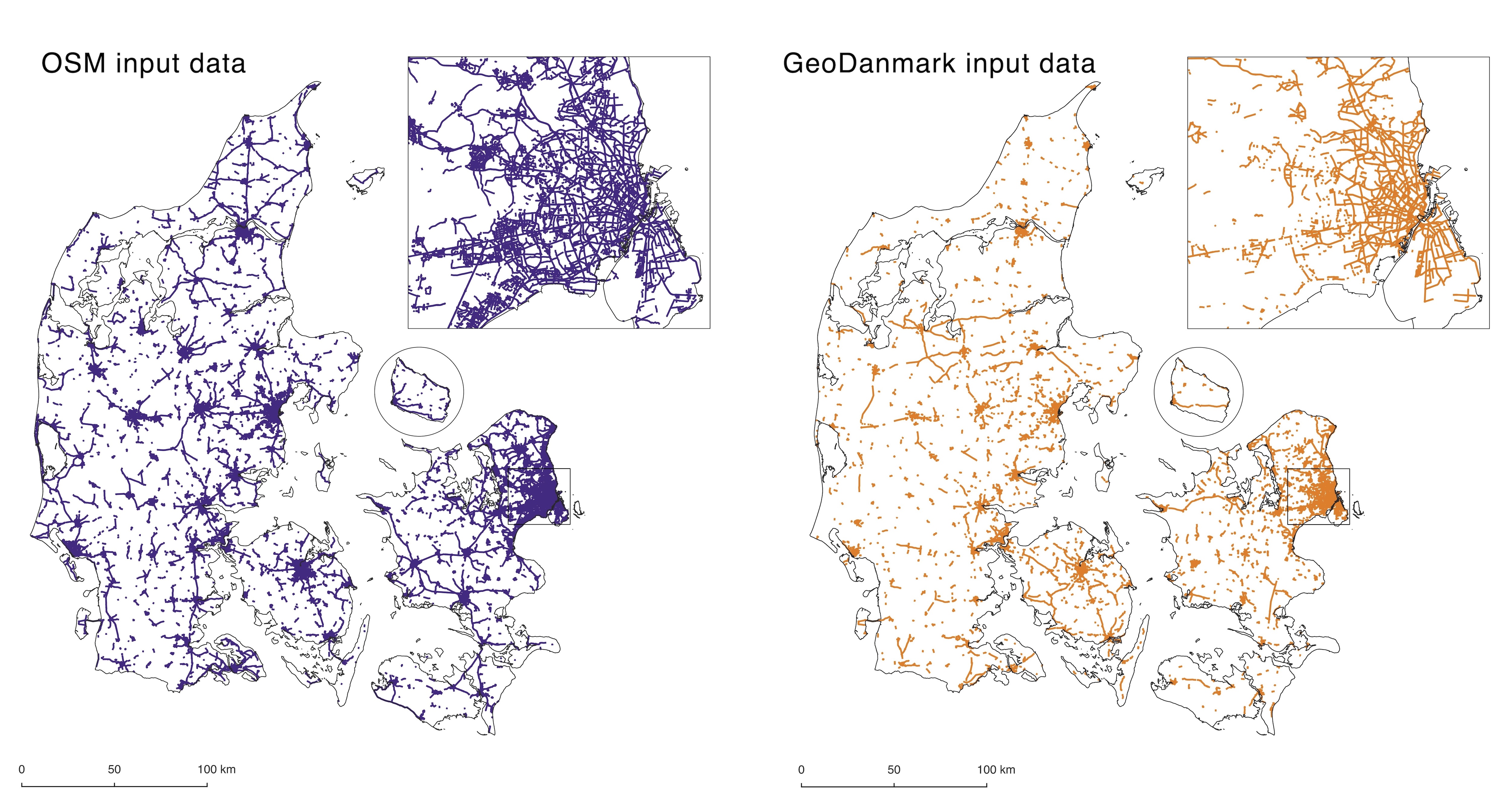}
\caption{\textbf{Overview of the two input data sets.} OSM bicycle infrastructure (left). GeoDanmark bicycle infrastructure (right). Copenhagen and surroundings in map insert.}
\label{fig:overview}
\end{figure}

To address the problem of bicycle infrastructure data with low or unknown quality, here we conduct a countrywide quality assessment of bicycle infrastructure data for the entire extent of Denmark. Our results are of particular interest for country-specific applications, while our methods are transferable to any other region or country. The two open data sets assessed in this study are the networks of dedicated bicycle infrastructure from the global collaborative mapping platform OpenStreetMap (OSM) \citep{openstreetmap_contributors_openstreetmap_2023} and from the national public data set GeoDanmark \citep{geodanmark_danmarks_2023} (Fig.~\ref{fig:overview}). We analyze the spatial data quality of bicycle infrastructure data for the entire country for both data sets, with special attention to network structure and spatial patterns in levels of data quality. In particular, we pose the following research question:

\smallskip
\noindent \emph{Is the spatial data quality of the OSM and GeoDanmark data sets adequate to support network-based analysis of cycling conditions in Denmark?}
\smallskip

\noindent To answer this question, we compare the two data sets through four data quality metrics: data completeness based on infrastructure density, data completeness based on feature matching, network structure, and OSM tag completeness. All metrics are computed with the Python-based open-source tool \texttt{BikeDNA} \citep{viero_bikedna_2023}. For each of the quality metrics, we then investigate spatial patterns, such as indications of spatial autocorrelation. This is, to our knowledge, the first investigation of spatial patterns of bicycle infrastructure data quality; the first study that assesses bicycle data quality for the entire country of Denmark; and one of the first studies to examine bicycle infrastructure data quality outside of urban areas. 

An assessment of the spatial data quality of a bicycle infrastructure data set, like the one presented here, has two main purposes: identifying and correcting specific data errors and informing data management with the goal of improving data quality. Our Denmark case study serves both purposes: first of all, we find substantial differences in not just the \textit{amount} of bicycle infrastructure, but also in \textit{where} bicycle infrastructure is mapped in the two data sets. Importantly, we conclude that the prevalent method for evaluating data completeness through density differences is inadequate for bicycle infrastructure data, due to the large variability in bicycle infrastructure mapping practices. Moreover, we find widespread topological errors, appearing for different reasons in the two data sets. These errors require customized solutions, in particular for any network-related purposes such as routing. Lastly, we find that the completeness of OSM tags relevant to bicycle conditions, such as road surface or street lighting, follow distinct spatial patterns, with large variations between the completeness of tags within compared to outside of urban centers.

The rest of the paper is organized as follows: first, we give a brief overview of previous work on spatial data and bicycle network data quality (Section~\ref{lit_review}), followed by an introduction to the data sets used in this study and data on bicycle infrastructure networks in general (Section~\ref{bicycle_data}). Next, we introduce the methods used for the spatial data quality evaluation (Section~\ref{methods}). We then present the results and what they can tell us about the bicycle infrastructure in Denmark (Section~\ref{results}). Lastly, we discuss potential applications, future work, and limitations of this study (Section~\ref{discussion}), and end with a conclusion that sums up our findings (Section~\ref{conclusion}).

\section{Literature review} 
\label{lit_review}

The quality of spatial data in general, and of volunteered geographic information (VGI) and other crowdsourced data sets in particular, is overall well-studied \citep{medeiros_solutions_2019, fonte_assessing_2017, degrossi_taxonomy_2018} -- but much less so for bicycle infrastructure data, for which very few studies exist. Therefore, we conduct our literature review in two steps: first, we review previous work on spatial data quality assessment, with a focus on OSM bicycle infrastructure data quality (Section~\ref{litrev-dq}). Then, we provide a general typology of bicycle infrastructure data and an overview of common data sources and error types (Section~\ref{common_issues}), which motivates the methods (Section~\ref{methods}) and their applicability to other locations than Denmark.

\subsection{Previous work on spatial data quality assessment}
\label{litrev-dq}
Spatial data quality encompasses both the quality of the \textit{spatial geometries} and the \textit{attributes and information} associated with each geometry \citep{fonte_assessing_2017}. An increasingly popular approach to spatial data quality, particularly of VGI and OSM data \citep{biljecki_quality_2023}, is the `fitness-for-purpose' concept, which asks whether a data set fulfills the requirements for a given \textit{use case}, instead of using a formalized definition of data quality \citep{devillers_towards_2007, brando_quality_2010, barron_comprehensive_2014, zhang_how_2015, brovelli_towards_2017}. This is the approach to spatial data quality we use in this study.

Studies on spatial data quality commonly distinguish between \textit{intrinsic} and \textit{extrinsic} methods for quality assessment. Intrinsic methods evaluate the internal properties of one single data set, while extrinsic methods compare the data set to an external (`reference') data set \citep{barron_comprehensive_2014}. In the case of OSM, studies using intrinsic methods mostly analyze edit history, contributors, or tag completeness \citep{kesler_tracking_2011, neis_comparison_2013,
barron_comprehensive_2014, grochenig_estimating_2014, hashemi_assessment_2015}, while studies based on extrinsic methods compare OSM data with other data sets from e.g.~administrative sources \citep{haklay_how_good_2010, koukoletsos_assessing_2012, neis_street_2012, gartner_is_2015, brovelli_towards_2017}. 

Regardless of the method used, most studies on OSM road network data quality agree on two points: first, OSM road network data are of a \textit{generally} high quality and completeness; and second, the quality of OSM road network data suffer from \textit{large spatial variations}. Data quality variations are seen across city-level, national, and international scales \citep{haklay_how_good_2010, neis_street_2012, brovelli_towards_2017, barrington-leigh_worlds_2017}. Moreover, these variations occur in the spatial distribution of added OSM tags \citep{almendros-jimenez_analyzing_2018, zhang_detecting_2021} and between different parts of the road network. For example, data on infrastructure for active mobility often lags behind data on infrastructure for motorized mobility \citep{neis_street_2012, neis_comparison_2013}. Finally, OSM data quality tends to be lower in less densely populated areas \citep{haklay_how_good_2010, barrington-leigh_worlds_2017}. Due to these heterogeneities in OSM data quality, many studies see the need to subdivide study areas to present the results on a local scale \citep{haklay_how_good_2010, forghani_quality_2014, brovelli_towards_2017}. 

So far, very few studies have investigated the quality of bicycle infrastructure data specifically. Notable examples are \citet{hochmair_assessing_2015} and \citet{ferster_using_2020}, which both assess the quality of OSM data on dedicated bicycle infrastructure in selected North American cities through extrinsic comparisons with reference data sets. \citet{hochmair_assessing_2015} compute and compare the aggregated density of OSM bicycle infrastructure data to a reference data set and manually inspect tagging and completeness errors. \citet{ferster_using_2020} also compute aggregate density and match corresponding features to identify overall differences and the exact locations where OSM and other open data sets disagree. Both studies conclude that, although OSM data are of a generally high quality and in many places in concordance with local reference data sets, there are substantial spatial differences in data completeness, mapping practices, and tagging precision within and between the examined cities. At the same time, the two studies draw opposing conclusions regarding different bicycle infrastructure types: \citet{hochmair_assessing_2015} conclude that OSM in the examined locations is more complete for protected than for unprotected bicycle infrastructure, while \citet{ferster_using_2020} find the biggest concordance between OSM and reference data precisely for unprotected bicycle infrastructure. These contradictory findings emphasize the need for local spatial data quality assessments of OSM data, particularly for  less frequently studied parts of the data set, such as bicycle infrastructure. 

To sum up, we can conclude that bicycle infrastructure data quality is subject to substantial spatial variations: error types vary between locations, and findings cannot be generalized from one location to another. Moreover, there are many differing tagging practices and data models for bicycle infrastructure, which makes it necessary to adapt spatial data quality assessment methods specifically to this type of data. Finally, the two identified studies on OSM bicycle infrastructure data quality are specific to the North American context and cover only the aspects of data completeness and misclassification, without addressing data topology. However, correct topology is required by many bicycle infrastructure data applications, such as routing and accessibility analysis. Thus, the challenge of determining the fitness-for-purpose of bicycle infrastructure data for these purposes has so far remained unaddressed. 

\subsection{Typology, data sources, and common quality issues of bicycle infrastructure data}
\label{common_issues}
\begin{figure}[h]
\centering
\includegraphics[width=0.8\textwidth]{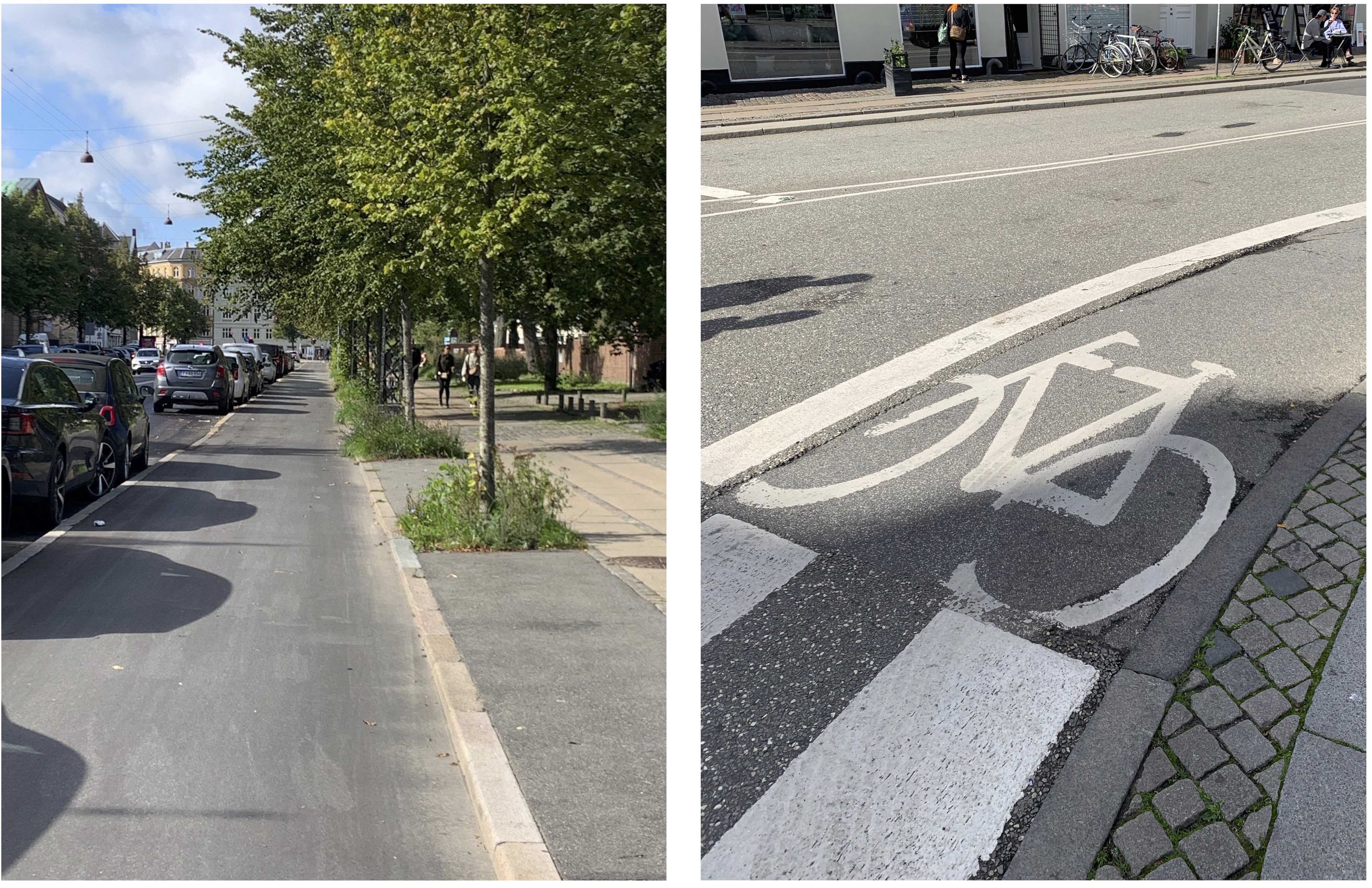}
\caption{\textbf{Types of bicycle infrastructure.} The analysis includes dedicated bicycle infrastructure, which can be either \emph{protected} bicycle tracks (left) or \emph{unprotected} bicycle lanes (right).}
\label{fig:bicycle_infra_illustration}
\end{figure}

Within the context of this article, we define `bicycle infrastructure' as the elements of the road and path network that are \textit{dedicated exclusively to cyclists}. As illustrated in Fig.~\ref{fig:bicycle_infra_illustration}, dedicated bicycle infrastructure can be either protected (bicycle tracks, physically separated from the motorized traffic) or unprotected (bicycle lanes, with no physical separation from the motorized traffic). Dedicated bicycle infrastructure is crucial to encourage cycling \citep{fosgerau_bikeability_2023}, and improves both the actual and the experienced cycling safety \citep{kamel_impact_2021, gossling_subjectively_2022}. However, data on where this infrastructure exists are often inadequate \citep{hochmair_assessing_2015, ferster_using_2020, ramboll_walking_2022}. At the same time, physical networks of dedicated bicycle infrastructure tend to be significantly more fragmented than networks for motorized traffic, and suffer from many missing links and disconnected components \citep{natera_orozco_data-driven_2020, vybornova_automated_2022, reggiani_multi-city_2023}. These two issues can therefore appear indistinguishable, creating a specific challenge for bicycle infrastructure.

There are two main sources for open data on bicycle infrastructure: online mapping platforms, where OSM is the most well-known and widely used \citep{ferster_using_2020, nelson_crowdsourced_2021}, and public agencies, such as municipal administrations or national mapping agencies. OSM is the go-to data source for research on the built environment and for projects and applications relying on open road network data \citep{carlino_b_2023, peopleforbikes_bna_2023, cyclestreets_cyclestreets_2023}, whereas administrative data from public agencies to our knowledge are primarily used for planning and administrative purposes. Regardless of the data source, data on bicycle infrastructure are often of an unknown, heterogeneous, or low quality \citep{hochmair_assessing_2015, ferster_using_2020, ramboll_walking_2022, hvingel_gode_2023}. For non-OSM data, such as administrative data, the lower quality of bicycle data has been explained with a lack of resources, as active mobility is given lower priority in contrast to motorized modes \citep{ramboll_walking_2022}.

Below, we give a brief overview of the most prominent quality issues for bicycle infrastructure data from a network research perspective. For a more detailed introduction to the different types of data quality problems in bicycle infrastructure data, see \citet{viero_bikedna_2023}.

The starting point for most quality assessments is data \textit{completeness}, which indicates whether all existing objects are represented in the data. Issues with data completeness can be divided into errors of \textit{omission} and \textit{commission}, referring to missing or excess data, respectively (see Fig.~\ref{fig:common_issues}). Data completeness is also affected by \textit{misclassifications} (see below). Crowdsourced data that are added gradually over time are especially prone to suffer from incompleteness during early stages of data collection \citep{neis_street_2012}.

The next step is typically to assess data \textit{consistency}, which, for bicycle infrastructure data, includes complete and correct \textit{classification}, data \textit{topology}, and the \textit{data model} used to map bicycle infrastructure. Misclassification issues appear e.g.~when unprotected bicycle infrastructure is classified as protected, or vice versa. Topology issues can arise due to missing nodes at intersections or because of undershoots, i.e.~when infrastructure geometries are slightly too short and therefore do not connect (Fig.~\ref{fig:common_issues}). Finally, although differing ways of mapping bicycle infrastructure are not errors in themselves, they can be a hindrance for comparing different data sets and pose problems if a chosen data model does not support the desired data application. For example, in the data sets from our study, OSM uses a combination of mapping bicycle infrastructure to the center line (where bicycle infrastructure running along a road is mapped by adding a tag to the road center line) \textit{and} mapping bicycle tracks with their own geometries. Meanwhile, in GeoDanmark, bicycle infrastructure is always mapped with separate geometries, regardless of the infrastructure type (Fig.~\ref{fig:common_issues}).    

\begin{figure}[H]
\centering
\includegraphics[width=0.999\textwidth]{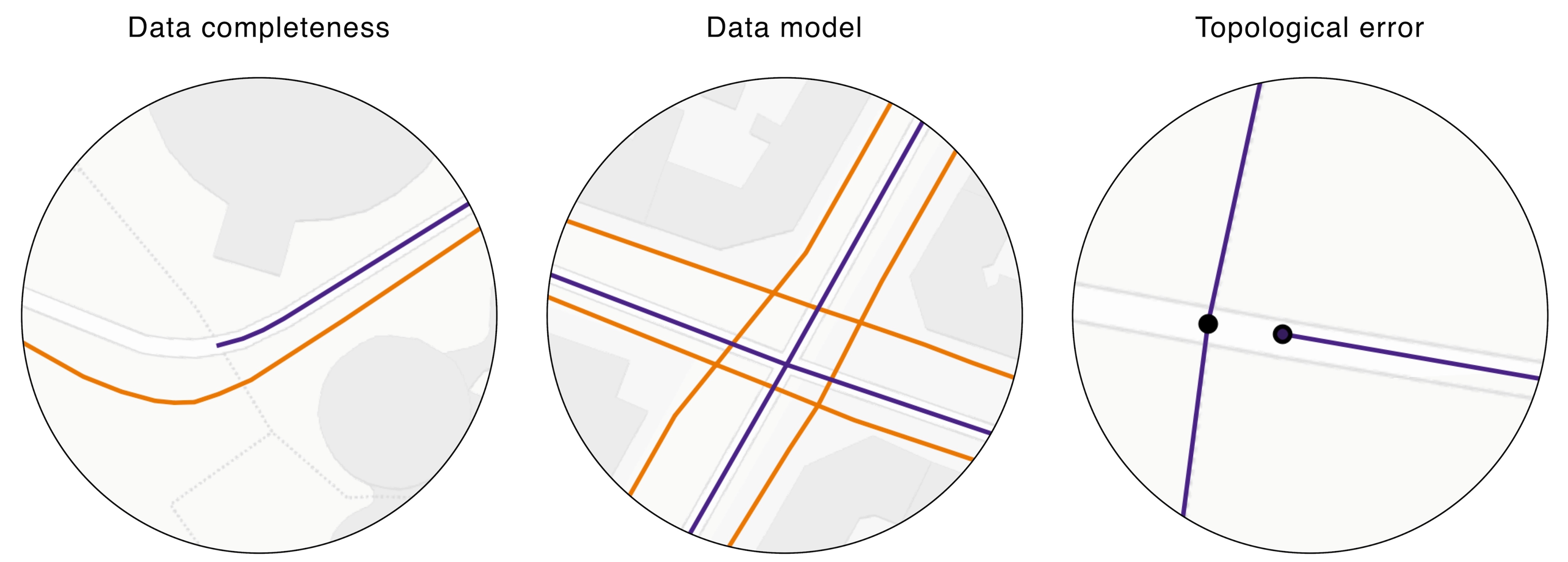}
\caption{\textbf{Common quality issues in bicycle infrastructure data.} Left: Different levels of data completeness, with an error of commission resulting in a longer bike path in the GeoDanmark data (orange) than OSM (purple). Center: Different data models in OSM (purple) and GeoDanmark (orange), with OSM using a center line mapping and GeoDanmark mapping all infrastructure with separate geometries. Right: Example of an undershoot in OSM data.}
\label{fig:common_issues}
\end{figure}

\section{Data} 
\label{bicycle_data}

The case study makes use of six different data sets: OSM and GeoDenmark as the two data sets on bicycle infrastructure to be analyzed; and four auxiliary data sets providing, respectively, a delimitation of the study area (the entire extent of Denmark), administrative subdivisions at the municipal level, municipal population sizes, and the local population density across the country.

The OSM data were downloaded from Geofabrik \citep{geofabrik_our_2020}. The GeoDanmark road network data is a national, open data set including the main road network, bicycle infrastructure along the car road network, and different types of paths. GeoDanmark data were downloaded from the national data portal Datafordeleren \citep{datafordeler_datafordelerdk_2023}. Both data sets were preprocessed with \texttt{BikeDNA} \citep{viero_bikedna_2023} using the Python libraries \texttt{pyrosm} \citep{tenkanen_htenkanenpyrosm_2021},  \texttt{OSMnx} \citep{boeing_osmnx_2017}, and \texttt{momepy} \citep{fleischmann_momepy_2019}. The OSM network contains \num{88997} network edges with a total length of \num{15333}~km and \num{90804} nodes. The GeoDanmark network contains \num{50856} network edges with a total length of \num{8676}~km and \num{51224} nodes. Both the OSM and the GeoDanmark data sets originally contain the full road network, but only the subsets with dedicated bicycle infrastructure are used in this analysis (see supplementary information for the queries used to extract the subsets from the full network).

The data sets on study area delimitation (geographical extent of Denmark with an area of \num{43057} \si{km^{2}}) and municipal names and boundaries have been downloaded from Datafordeleren \citep{datafordeler_datafordelerdk_2023}. Finally, we obtained municipality population sizes from Statistics Denmark \citep{statistics_denmark_statistikbanken_2023} and a population density raster from the European Commission's Global Human Settlement Layer \citep{schiavina_ghs-pop_2023}. 

\section{Methods} 
\label{methods}

\begin{figure}[t]
\centering
\includegraphics[width=0.999\textwidth]{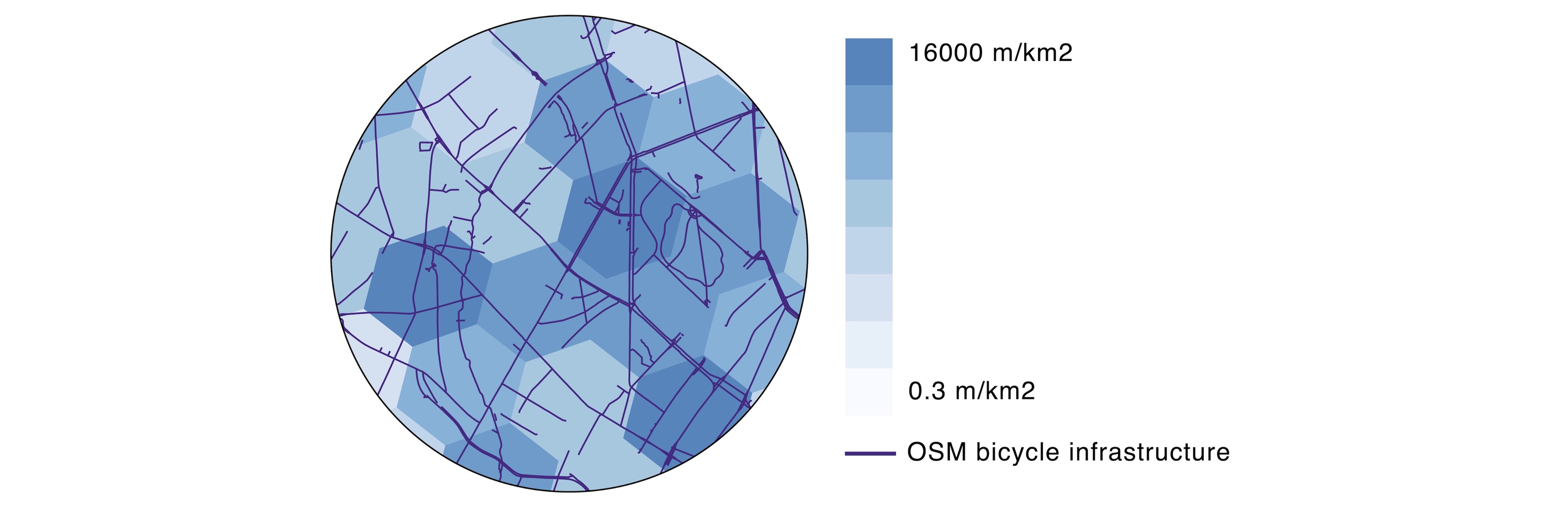}
\caption{\textbf{Example of hex grid aggregation.} Hex grid cells used to compute the local infrastructure density of OSM data.}
\label{fig:hex_illu}
\end{figure}

Below, we present the methods used in this study. The main idea of our approach is to first assess four metrics of data quality (data completeness based on infrastructure density, data completeness based on feature matching, network structure, and OSM tags), and second, to examine spatial patterns in all four of them. The four data quality metrics are computed with the previously developed tool \texttt{BikeDNA} \citep{viero_bikedna_2023}, in a version adapted to large data sets for the purpose of this study (\url{https://github.com/anerv/BikeDNA_BIG}). We compute quality metrics both globally, i.e.~for the whole data set, and locally, i.e.~for each cell in a H3-based hexagonal grid \citep{uber_h3-py_2023} covering the study area (see Fig.~\ref{fig:hex_illu}). The grid cell size is configurable in our workflow; here we set it to resolution \num{8} with an average cell size of \num{0.74} \si{km^{2}} (see the supplementary information for other parameter values used in the analysis). In the second methodological step, spatial patterns in data quality for all four metrics are analyzed through spatial autocorrelation \citep{anselin_local_1995, getis_reflections_2007} and by examining the spatial correlation of quality metrics with population densities (\url{https://github.com/anerv/bikedna_dk_analysis}).

\subsection{Data completeness from infrastructure densities}

Computing differences in infrastructure density is a common and computationally cheap way of assessing differences in data completeness \citep{haklay_how_good_2010, neis_street_2012}. We examine data completeness based on infrastructure density in three steps: first, comparing the total lengths of the OSM and GeoDanmark data; second, comparing infrastructure densities at the municipal level; and third, comparing infrastructure densities at the grid cell level. To compensate for the different data models in OSM and GeoDanmark data (Fig.~\ref{fig:common_issues}), our computation of data completeness is not based on the length of the geometries in each data set, but instead uses the concept of \emph{infrastructure length}, by which we consider information on allowed cycling direction and mapping approaches. For example, an OSM center line mapping of bicycle lanes on both sides of a \num{100} meter long road will be counted as 200 meters to allow for comparison with the data model in GeoDanmark, where the same infrastructure would be mapped with two separate \num{100} meter geometries on each side of the road. 

While varying levels of local infrastructure density can indicate areas with missing or surplus data, they do not reveal specific omission or commission errors. The local measures of infrastructure density are moreover aggregate measures and thus inherently prone to issues, such as hiding or exaggerating spatial differences. For this reason, we also apply a feature matching algorithm \citep{koukoletsos_automated_2011, koukoletsos_assessing_2012, will_development_2014, viero_bikedna_2023} on the two data sets to identify corresponding objects in OSM and GeoDanmark data.

\subsection{Data completeness from feature matching} \label{methods_fm}

Feature matching (i.e. the identification of corresponding features in different data sets) is the most exact method for obtaining differences in data completeness. The feature matching procedure divides the geometries in both data sets into segments of equal length and uses a combination of configurable maximum thresholds for segment distance, Hausdorff distance, and the angle between segments from the corresponding data sets to find the best match \citep{koukoletsos_automated_2011, koukoletsos_assessing_2012, will_development_2014}. Just as for the comparison of data completeness, we compute results both globally (as aggregate values for the entire data set) and locally (as the total count and percentage of matched and unmatched segments in each grid cell).

\subsection{Network structure}

Consistency in network structure is crucial for network-based applications like routing or accessibility analysis, which are made impossible by topological errors and inconsistent network fragmentation. An analysis of fragmentation and topological errors is, as a starting point, an intrinsic evaluation aimed at detecting internal (in)consistencies. Nevertheless, due to the scattered nature of many actual bicycle networks, a comparison of network fragmentation and topological errors between two data sets is a useful, and sometimes necessary, method for distinguishing between poor \textit{data} quality and poor \textit{network} quality without conducting a manual validation. Here, we focus on two aspects of network structure: \textit{network components}, as a proxy for the fragmentation or connectivity in the data, and \textit{undershoots} (see Section~\ref{common_issues}), as an example of topological errors.

A disconnected network component is a subset of a network where all nodes of the component can reach each other internally, but no nodes of the component can reach the rest of the network. Most actual bicycle infrastructure networks are made up from many disconnected components \citep{furth_network_2016, natera_orozco_data-driven_2020, szell_growing_2022}. Data on networks of bicycle infrastructure will thus often be correctly divided into many disconnected components, but can at the same time suffer from an additional fragmentation due to data quality issues. Unwanted components can for example occur due to incomplete data resulting in `missing links' \citep{vybornova_automated_2022}, snapping issues, and imprecise geographic coordinates resulting in overlapping but not truly connected network edges. In the case of OSM or similar data sets, a missing bicycle tag on part of a road can turn a single piece of infrastructure into two disconnected fragments in the data (see Fig.~\ref{fig:detected_undershoots}). For bicycle infrastructure data created without considering routing, disconnected components can also appear if e.g.~bike lanes are not explicitly connected across intersections. While most disconnected bicycle components are connected by the road network, these connections do not serve cyclists who are unable or unwilling to bike in mixed traffic. Therefore, the connectivity of the network of dedicated bicycle infrastructure is of primary importance for e.g.~analysis of low stress cycling and accessibility \citep{mekuria_low-stress_2012, lowry_quantifying_2017, reggiani_understanding_2021}.

By running \texttt{BikeDNA}, we acquire the number of network components in each data set, the distribution of infrastructure length per component, the local component count (how many different disconnected components a grid cell intersects), and the local and global numbers of undershoots. For further analysis and visualization, this step also indicates which network nodes have been identified as undershoots, and which component a given network edge belongs to. In this study, we define undershoots as dangling nodes within three meters of a network edge to which the node is not connected (Fig.~\ref{fig:common_issues}).

\subsection{OSM tags}

OSM tags are key-value pairs which define the core feature type of mapped geometries, as well as any additional information. OSM uses an open tagging system with best practices, but no enforcement of standards \citep{hochmair_assessing_2015}. Correct tagging is thus a crucial first step towards OSM bicycle infrastructure data completeness. Incorrect tagging leads to errors of omission and commission for local bicycle networks in OSM \citep{hochmair_assessing_2015, ferster_using_2020}. Apart from the tags defining the presence of bicycle infrastructure, additional information on the infrastructure, such as surface and lighting conditions, is also of relevance to many bicycle planning and research projects \citep{wasserman_evaluating_2019}. The completeness and consistency of OSM tags can be used as an indicator of spatial data quality \citep{mooney_annotation_2012, almendros-jimenez_analyzing_2018, biljecki_quality_2023}. Previous research has found large spatial variations in tagging patterns and completeness, both for the road network data and other features mapped in OSM \citep{mooney_annotation_2012, barron_comprehensive_2014, hochmair_assessing_2015, almendros-jimenez_analyzing_2018,  biljecki_quality_2023}.

To examine the extent to which the OSM data in the study area contain information relevant to e.g.~a mapping of Levels of Traffic Stress \citep{mekuria_low-stress_2012} and bikeability, we analyze the local share of bicycle infrastructure with tags for four different attributes: 

\begin{itemize}
    \item Infrastructure surface: `surface'/`cycleway:surface'.
    \item Presence of street lights: `lit'.
    \item Width of the infrastructure: `width'/`cycleway:width'.
    \item Speed limit for motorized traffic: `maxspeed'.
\end{itemize}

Due to the absence of ground truth data, our analysis only considers the existence, but not the correctness of tags.

\subsection{Spatial patterns in local data quality metrics} 
\label{methods_evaluating}

Identifying areas with particularly low or high data quality can help understand why quality issues occur. Our goal is to establish whether there is a discernible spatial pattern in data quality or whether variations in data quality are randomly distributed. To this end, we use spatial autocorrelation to identify areas with particularly low or high data quality for results for infrastructure density, feature matching and OSM tag completeness. Results for differences in infrastructure density are moreover examined at the municipal level to examine the role of differing data maintainers. GeoDanmark data are collected on national level, but the Danish municipalities play a central role in data maintenance and updates \citep{geodanmark_produktion_2020}. We are thus interested in detecting whether the municipal involvement in data maintenance is reflected in the data quality, especially since there are indications that the municipalities are following different mapping practices for the classification of bicycle infrastructure \citep{hvingel_gode_2023}. For OSM, we do not expect the data quality to follow municipal boundaries -- unless the local administrations contribute to OSM, of which there are some examples internationally \citep{osm_openstreetmap_2022}. To detect discrepancies between GeoDanmark and OSM, we aggregate quality metrics for both data sets at the municipal level. 

An analysis of spatial patterns in data quality at the municipal level can however not stand alone, since the level of spatial aggregation makes it vulnerable to the modifiable areal unit problem (MAUP), which describes how spatial aggregation can distort, conceal or exaggerate spatial patterns in the data \citep{mennis_problems_2019}. To avoid this pitfall, we supplement the analysis of differences in infrastructure length at the municipal level with a further analysis of spatial autocorrelation in differences in data completeness, feature matching success, and OSM tag completeness at the grid cell level, which has a much higher spatial resolution.

Spatial autocorrelation describes how the value of a variable varies across space by quantifying to what extent data points in close proximity have similar values \citep{wu_global_2019}. In this study, the global spatial autocorrelation is measured using Moran's I. Values for Moran's I range from \num{-1} to \num{1}, where values above \num{0} indicate a clustering of \emph{similar} values. Values below \num{0} indicate that data points tend to be close to \emph{dissimilar} values, while a Moran's I close to \num{0} indicates spatial randomness \citep{rey_geographic_2020}. Since we are interested not only in the degree of spatial autocorrelation, but also in the exact spatial location of potential clusters of similar values, we also calculate Local Indicators of Spatial Association (LISA) \citep{anselin_local_1995} using local Moran's I. We calculate both global and local Moran's I with the Python library \texttt{ESDA PySAL} \citep{rey_pysal_2007}. All reported clusters of local spatial autocorrelation are significant at a pseudo p-value of \num{0.05}.

Computing spatial autocorrelation on a fragmented network with a highly uneven network density is an analytical challenge, since there is no obvious way of defining the spatial weight matrix on which the computation is based. For this reason, we compute the spatial autocorrelation for results aggregated at a local grid cell level, using a hexagonal grid with a row-standardized spatial weight matrix based on the k-nearest neighboring grid cells, with k=6. To check the sensitivity of the results to the definition of the spatial weights, we repeated a part of the analysis with spatial weights based on \num{12} and \num{18} nearest neighbors and with distance bands of \num{1000} and \num{2000} meters, and found that changing those parameters did not alter the general patterns of spatial autocorrelation (Table S2, S3, and S4 in the supplementary information).

To explore how different methods for data completeness evaluations perform, we furthermore compare differences in local infrastructure density with the results from the more exact, but also more computationally expensive, feature matching of corresponding network segments. Finally, to establish whether the common link between OSM data quality and population densities also holds for Danish bicycle infrastructure data, we analyze how infrastructure density differences and completeness of OSM tags correlate with local population density.

\section{Results} 
\label{results}

Below, we organize our main findings into four subsections, one for each of the four quality metrics (data completeness based on infrastructure densities, feature matching, network structure, and OSM tags). Due to the size of the input data (more than \num{8600} kilometres of bicycle infrastructure in the smaller GeoDanmark data set) and the study area (\num{43057} \si{km^{2}}), it is not feasible to present results for all parts of the network in detail. Instead, we highlight the most relevant findings from the perspective of a bicycle network analysis. In addition, we emphasize findings that are of relevance not only to bicycle infrastructure data in Denmark, but additionally help us understand the quality and characteristics of bicycle infrastructure data in general. Results are aggregated, examined, and presented at two scales: at the global (study area) and at the local (grid cell) level. Results for differences in infrastructure density are additionally aggregated at the municipal level to explore the influence on data completeness from differing data maintainers.

\begin{table}[H]
\centering
\begin{tabular}{|l|r|r|}
\hline
\multicolumn{3}{|c|}{\textbf{Extrinsic Quality Comparison}} \\
\hline
\textbf{Metric}& \textbf{OSM} &\textbf{GeoDanmark} \\
\hline
Total infrastructure length (km) & \num{20681} & \num{8676} \\
Protected infrastructure length (km) & \num{17876} & \num{4264} \\
Unprotected infrastructure length (km) & \num{2804} & \num{4412} \\
Nodes & \num{90804} & \num{51224} \\
Dangling nodes & \num{46426} & \num{11218} \\
Undershoots & \num{157} & \num{339} \\
Components & \num{10686} & \num{4408} \\
Length of largest component (km) & \num{3433} & \num{1018} \\
Largest component's share of network length & \num{22}{\%} & \num{12}{\%} \\
\hline
\end{tabular}
\smallskip
\caption{\textbf{Extrinsic summary}. Selected metrics from extrinsic comparison of OSM and GeoDanmark data for all of Denmark.}
\label{table:extrinsic_table}
\end{table}

\subsection{Results for data completeness: Infrastructure densities}
\label{completeness_section}

\begin{figure}[H]
\centering
\includegraphics[width=0.999\textwidth]{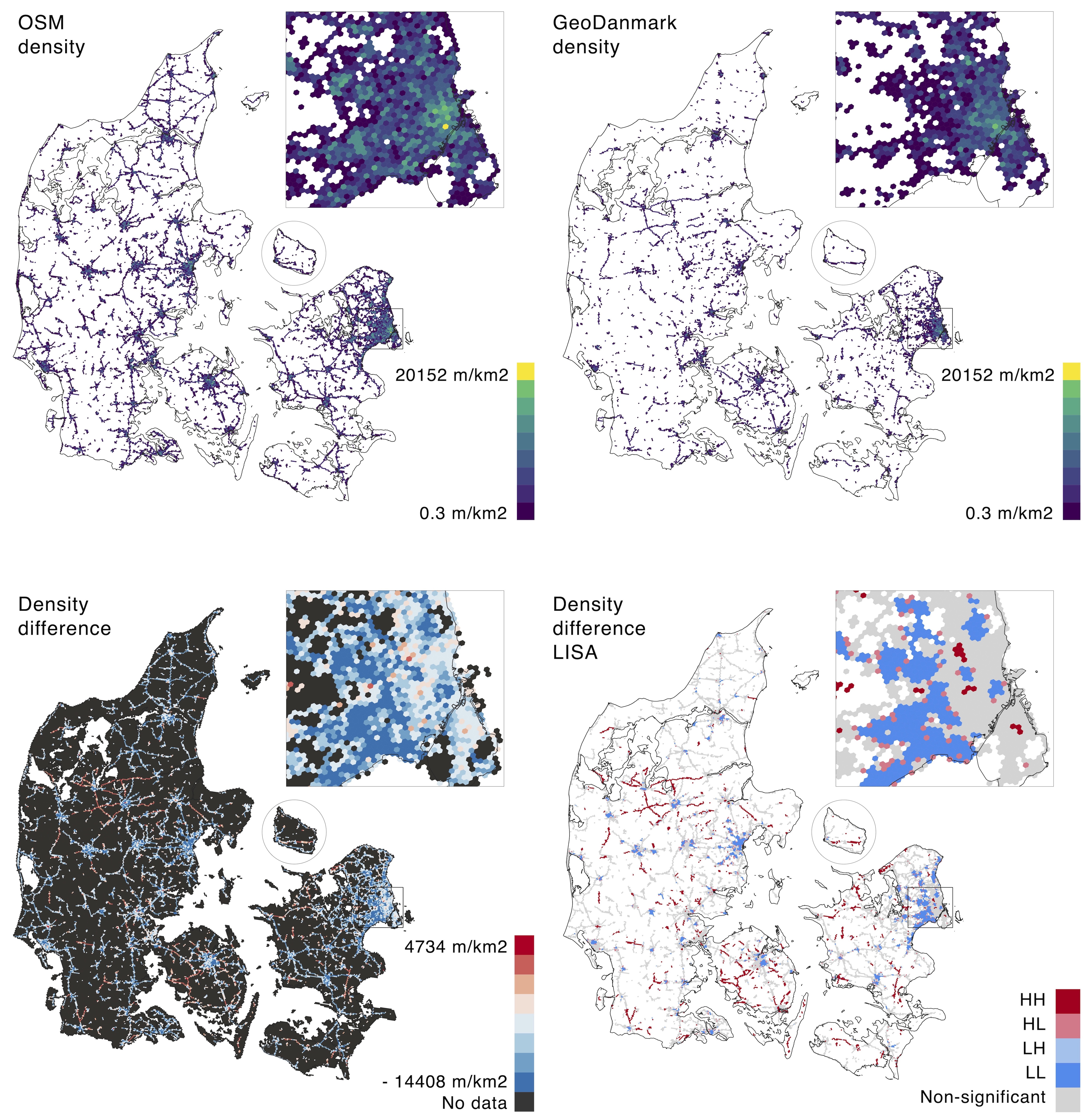}
\caption{\textbf{Infrastructure density at the grid cell level}. Top left: Bicycle infrastructure density from OSM data. Top right: Bicycle infrastructure density from GeoDanmark data. Bottom left: Difference in infrastructure density between OSM and GeoDanmark data. Areas with negative values (blues) have a higher density of OSM data. Areas with positive values (reds) have a higher density of GeoDanmark data. Bottom right: Analysis of local spatial autocorrelation of infrastructure density differences using Moran's I. Red areas, or 'High-High' (HH), indicate significant clusters of high values (p<\num{0.05}). Blue areas, or 'Low-Low' (LL), indicate significant clusters of low values. 'High-Low' (HL) represent high values surrounded by low values, while 'Low-High' (LH) represent low values surrounded by high values. In this context, a HL area means high relative GeoDanmark density surrounded by high relative OSM density, while LH represents high relative OSM density surrounded by high relative GeoDanmark density. Map inserts shows Copenhagen and surroundings in a bigger scale.}
\label{fig:density_diff}
\end{figure}

Comparing the total infrastructure length in both data sets at the global level (Table~\ref{table:extrinsic_table}), the OSM data set contains more than twice as much bicycle infrastructure as GeoDanmark. From a comparison of the different categories of bicycle infrastructure (protected vs.~unprotected), it is clear that this large difference is mostly attributable to the mapping of protected infrastructure. While GeoDanmark contains almost equal lengths of unprotected and protected infrastructure (\num{4412} \si{km} and \num{4264} \si{km} respectively), the OSM data only contain approximately \num{2784} \si{km} of unprotected, but \num{17856} \si{km} of protected bicycle infrastructure. The local differences in infrastructure density (Fig.~\ref{fig:density_diff}) moreover reveal substantial local variation in data completeness between the two data sets: out of the \num{16064} grid cells with data from either or both data sources, \num{81}{\%} of the cells have more bicycle infrastructure mapped in OSM, \num{19}{\%} have more in GeoDanmark.

At the municipal level, \num{96} out of \num{98} municipalities have more data in OSM than in GeoDanmark, with relative infrastructure length differences ranging between \num{2}{\%} and \num{94}{\%} (see Table S5). The differences in infrastructure length thus vary greatly between the Danish municipalities, leading to both large discrepancies between the sum of a municipality's total infrastructure and its ranking based on infrastructure length between the two data sets. For example, Aarhus municipality is the municipality with most bicycle infrastructure data in OSM, but is only in fourth place when ranking municipalities by infrastructure length in GeoDanmark data (see Fig.~S10). 

\begin{figure}[H]
\centering
\includegraphics[width=0.999\textwidth]{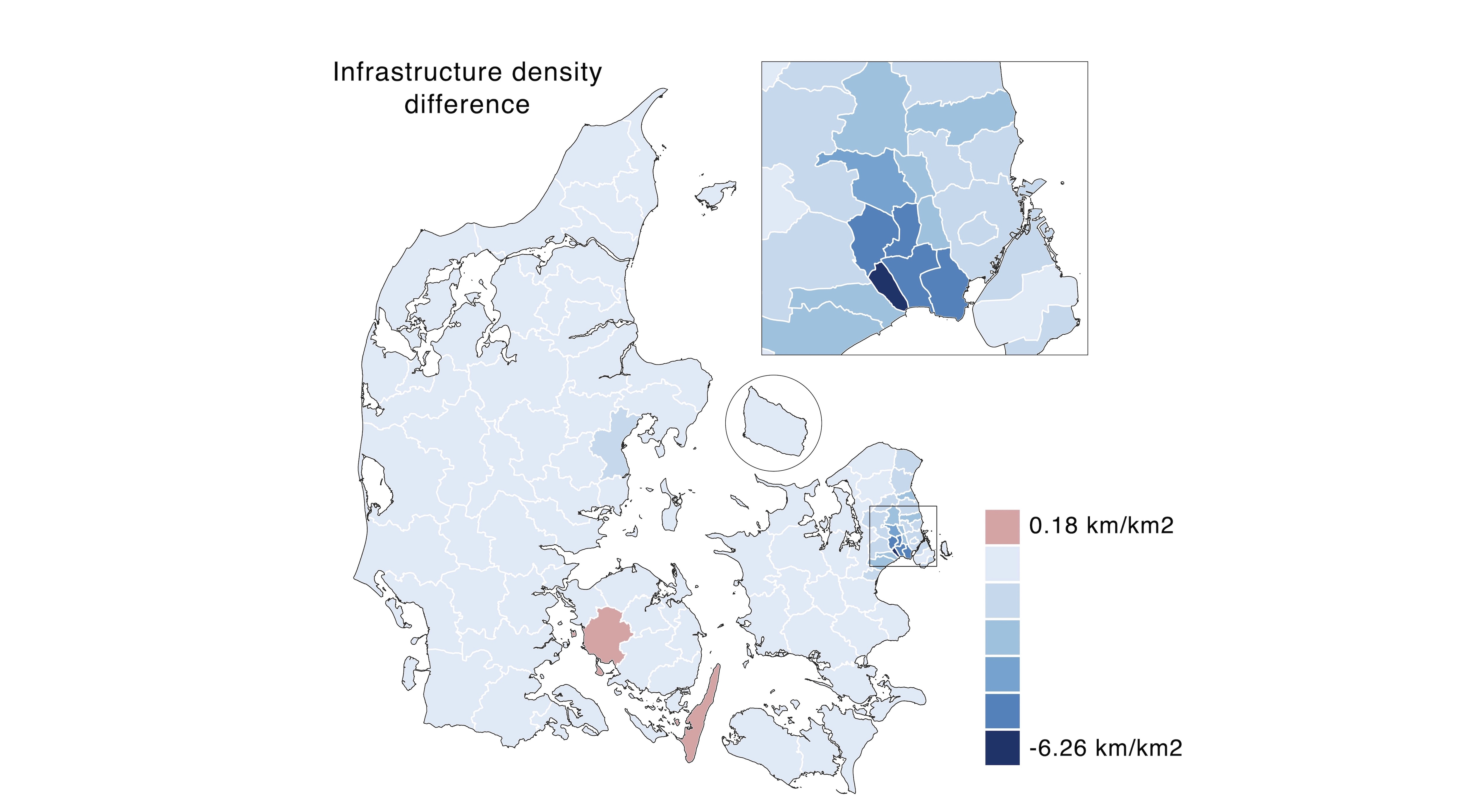}
\caption{\textbf{Difference in infrastructure density between OSM and GeoDanmark data at the municipal level}. The infrastructure density difference is computed as \textit{GeoDanmark \si{km/km^{2}}} - \textit{OSM \si{km/km^{2}}}. Negative values (blue) indicate municipalities where OSM data have a higher density; positive values (red) indicate municipalities where GeoDanmark data have a higher density. Out of the 98 municipalities, only two have more infrastructure mapped in GeoDanmark than OSM.}
\label{fig:muni_dens_diff}
\end{figure}

At the grid cell level, the differences in infrastructure density (Fig.~\ref{fig:density_diff}, bottom left) show a pattern of equal infrastructure density (light areas) or higher densities of OSM data (blue areas) in urban cores. Other parts of the country show no clear trend, with a combination of areas with more data in OSM and areas with more data in GeoDanmark (red areas). From the insert of Copenhagen, it is however also clear that there are local exceptions to this tendency. A test for correlation between infrastructure density differences and population density indicates some correlation, with more highly populated areas having more data in OSM, but also many exceptions to this trend (Fig.~S2). 

To statistically confirm that there are clusters of high and low density differences, we analyze the local values of infrastructure density differences for spatial autocorrelation. We use Moran's I to test for any global spatial clustering and Local Moran's I to identify specific clusters of similar or dissimilar values \citep{anselin_local_1995, rey_geographic_2020}. A situation of perfect spatial clustering would result in a Moran's I value of \num{1}, while a value of \num{0} would indicate a random pattern. Applying the spatial weight matrix based on k-nearest neighbors with k = \num{6}, as introduced in Section~\ref{methods_evaluating}, the global Moran's I statistic for infrastructure density differences is \num{0.46}, with a pseudo p-value of \num{0.001}. The results thus indicate a significant and positive, but not exceptionally strong clustering of similar values of infrastructure density differences (Fig.~S4). This numerical result is also visualized in Fig.~\ref{fig:density_diff}, bottom right: clusters of positive spatial autocorrelation of higher values of OSM infrastructure density (blue) appear in and around the major towns in Denmark, smaller clusters of higher values of both GeoDanmark (red) and OSM data are scattered across the country, while many areas show no statistically significant clustering (grey).

From these findings on data completeness at various levels of aggregation, we draw two conclusions. First, the spatial patterns in infrastructure density differences are not adequately captured at the municipal level, and a higher spatial resolution is required to show where differences occur. Second, the large variation in relative differences between GeoDanmark and OSM data completeness at the municipal level nevertheless suggests that there are differing municipal mapping practices for GeoDanmark data.

Due to incompatible classifications it is not possible to obtain exactly corresponding subsets of bicycle infrastructure: GeoDanmark data, per specification, only include bicycle infrastructure running along a road with motorized traffic, while there is no feasible way of just obtaining OSM bicycle infrastructure that runs in parallel with the car road network based on the OSM tags alone. This might explain some of the discrepancies between the total amount of \textit{protected} infrastructure in the two data sets. Moreover, as \citet{hvingel_gode_2023} have pointed out, bicycle infrastructure might be under-reported in GeoDanmark due to imprecise labelling with bicycle tracks being classified as a `main path' instead of the more specific `bicycle track'. However, this does not explain why OSM has fewer unprotected bicycle lanes than GeoDanmark, and suggests that variations in data completeness are more than just a misclassification issue. 

The lack of ground truth data makes any statements on the actual data completeness difficult. Manual inspections reveal errors of both omission and commission in both data sets, but without ground truth data, the extent to which discrepancies are due to missing or surplus data is unknown. Although some of the discrepancies in data completeness can be explained with different tagging and labelling conventions, the extent and spatial patterns in differences in data completeness suggest that OSM and GeoDanmark differ in both qualitative and quantitative terms when it comes to mapping of bicycle infrastructure. In the next section, we examine the results from feature matching to identify exactly where these differences occur.

\clearpage

\subsection{Results for data completeness: Feature matching}

In order to precisely detect where and to what extent two data sets are in agreement, their exact overlap needs to be identified. We achieve this with the feature matching method described in Section~\ref{methods_fm}. Here, we present the results of the initial feature matching performed with \texttt{BikeDNA} and the subsequent analysis of spatial autocorrelation of the results.
 
\begin{figure}[H]
\centering
\includegraphics[width=0.999\textwidth]{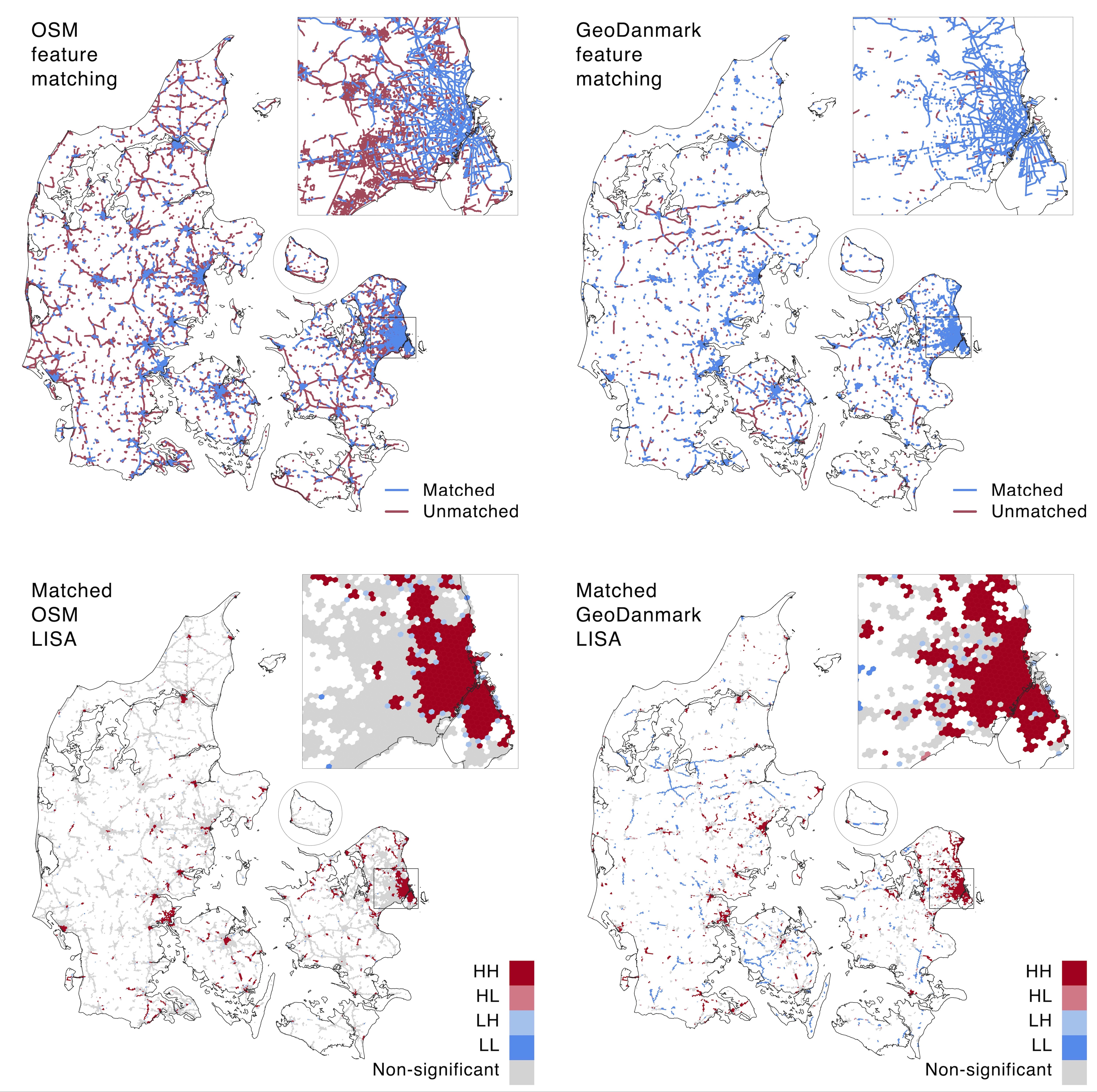}
\caption{\textbf{Feature matching results}. Top row: Matched (red) and unmatched (blue) segments in OSM data (left) and GeoDanmark data (right). Bottom row: Local spatial autocorrelation clusters of feature matching success (\% matched edges) for OSM data (left) and GeoDanmark data (right). The matching rates for both data sets are highest in the larger towns and cities and lowest in less densely populated areas.}
\label{fig:fm_all}
\end{figure}

A naive summing up of bicycle infrastructure lengths in each data set (disregarding differing mapping practices and data models) returns \num{15333} km of bicycle infrastructure in OSM -- almost twice as much as GeoDanmark with \num{8676} km. Nevertheless, only 64\% of the GeoDanmark segments match with an OSM segment, and only 23\% of OSM segments match with a GeoDanmark segment (Table~\ref{table:fm_table}). We provide a detailed illustration of feature matching results through a web map at \url{https://anerv.github.io/bikedna_webmap}. A visual inspection of the results confirms that the matching procedure is successful in most cases. The high levels of unmatched features in both data sets are therefore mostly explained by the two data sets containing not just \emph{different amounts} of bicycle infrastructure, but also bicycle infrastructure in \emph{different locations}. The spatial distribution of matched and unmatched segments (see Fig.~\ref{fig:fm_all}) thus suggest an even higher discrepancy between the two data sets than initially indicated by the differences in infrastructure density alone.

\begin{table}[b]
    \centering
    \begin{tabular}{|l|r|r|}
    \hline
    \multicolumn{3}{|c|}{\textbf{Feature Matching Results}} \\
    \hline
    \textbf{Metric}& \textbf{OSM} &\textbf{GeoDanmark} \\
    \hline
    Count of matched segments & \num{351476} & \num{564661} \\
    Length of matched segments (km) & \num{3490} & \num{5564} \\
    Percent matched segments & \num{23}{\%} & \num{64}{\%} \\
    Local min. of \% matched segments & \num{0}{\%} & \num{1}{\%} \\
    Local max. of \% matched segments & \num{100}{\%} & \num{100}{\%} \\
    Local average of \% matched segments & \num{53}{\%} & \num{83}{\%} \\
    \hline
    \end{tabular}
    \smallskip
    \caption{\textbf{Feature matching summary}. Selected results from matching of corresponding segments in OSM and GeoDanmark data.}
    \label{table:fm_table}
\end{table}
 
Global and local spatial autocorrelation of the local percentage of matched segments in OSM and GeoDanmark reveals a statistically significant positive spatial autocorrelation (pseudo p-value = \num{0.001}), with a Moran's I for the share of matched OSM and GeoDanmark segments of \num{0.48} and \num{0.52}, respectively. While the positive values for global spatial autocorrelation also cover large areas with no significant clustering of similar values (Fig.~S5), there are clear clusters of high matching success, especially around urban centers (Fig.~\ref{fig:fm_all}). 

\begin{figure}[h]
\centering
\includegraphics[width=0.999\textwidth]{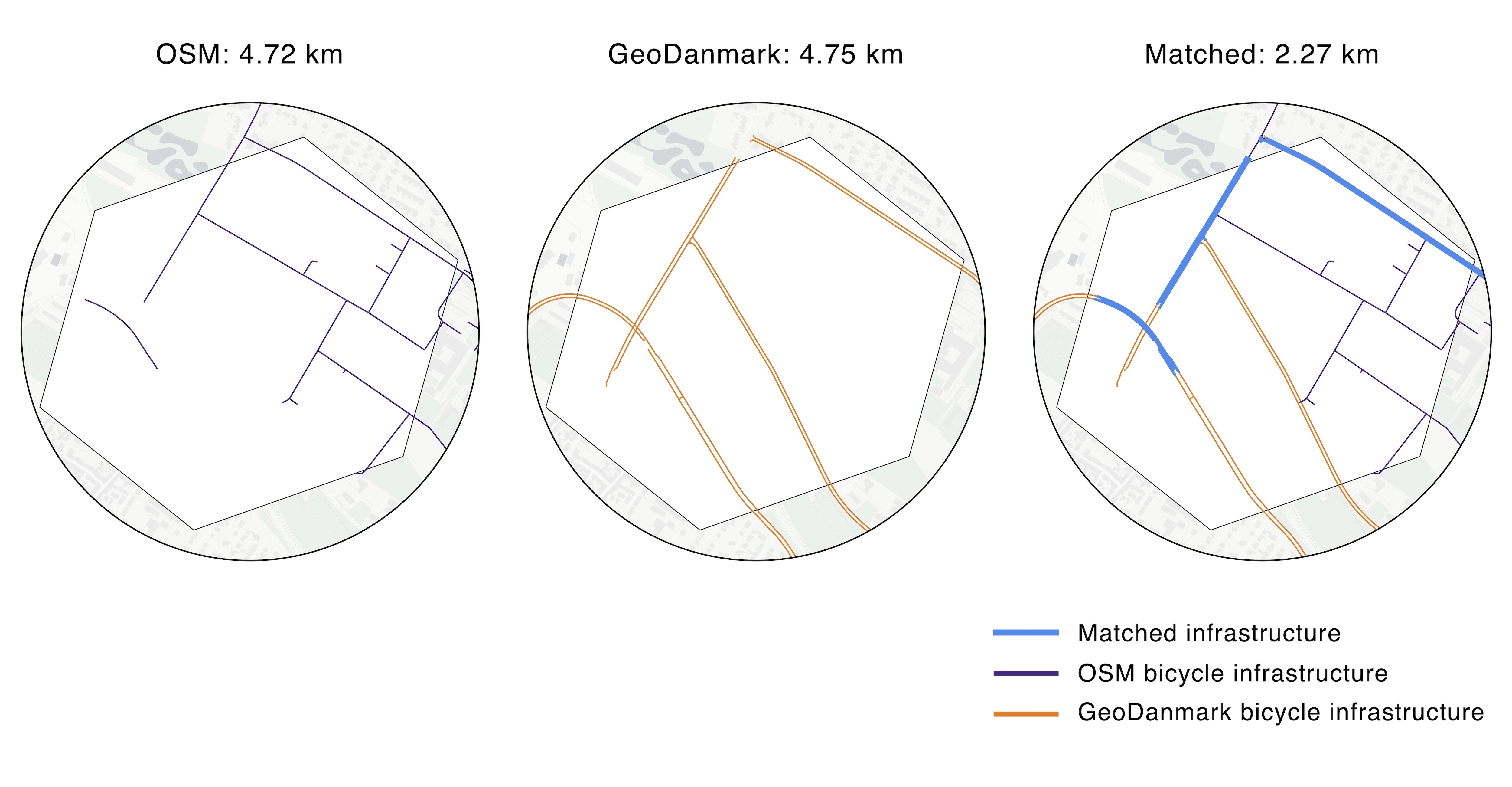}
\caption{An example of a grid cell (in white) with very low difference in the length of infrastructure between the two data sets, but high rates of unmatched OSM and GeoDanmark data. Left: OSM data. Center: GeoDanmark data. Right: Both data sets, matched data in blue. Despite having almost the same length of bicycle infrastructure, the two data sets barely overlap.}
\label{fig:low_dens_diff}
\end{figure}

The correlation between the local length of unmatched segments and the differences in infrastructure density can reveal whether unmatched data occurs due to a lack of data in the other data set. For OSM data, the correlation between the local length of unmatched segments and infrastructure density differences follow an expected pattern, with more unmatched OSM segments in areas where OSM contains more data than GeoDanmark (Fig.~S1). For GeoDanmark, this pattern is much less consistent: we see many locations with equal or higher amounts of data in OSM, but still high rates of unmatched GeoDanmark segments (Fig.~S1). In these locations, a low matching rate for GeoDanmark data can thus not be explained with GeoDanmark simply being more complete. On the contrary, a visual analysis confirms that the low matching rates are explained by the mapped infrastructure in the two data sets being complementary, and thus barely overlapping. The locations where infrastructure density differences are low, but rates of unmatched segments are high, illustrate how comparisons of infrastructure density can mask substantial differences in the actual bicycle infrastructure contained in different data sets (Fig.~\ref{fig:low_dens_diff}).

\clearpage

Summing up the assessment of data completeness from both infrastructure density differences and feature matching, it is not possible to provide a conclusive answer to how much bicycle infrastructure there is in Denmark. OSM contains substantially more data than GeoDanmark, which could indicate that the OSM data set is more complete. However, the large discrepancies in \emph{where} the two data sets include bicycle infrastructure, as well as the very different ratios of protected to unprotected infrastructure, suggest that either OSM still is missing a lot of data, that GeoDanmark data suffer from many errors of commission, or both. As both the completeness of OSM and GeoDanmark data are unknown and the size of the study area makes manual verification with e.g.~street view images unfeasible, we cannot conclude with certainty whether differences are due to errors of omission or commission. We are, however, able to identify where and to what extent differences exist. Notably, the concordance between the data sets is bigger in more densely populated, urban areas. Although there is no clear correlation between data concordance and population density (Fig.~S3), the high matching rates for denser urban areas are in line with previous research, which found data completeness to be higher in more densely populated areas \citep{fonte_assessing_2017, barrington-leigh_worlds_2017}. 

\clearpage

\subsection{Results for network structure} \label{structure_results}

To assess data quality from a network perspective, we examine two aspects of network structure: network fragmentation, which we measure by counting the disconnected network components; and the number of topological errors, which we measure by identifying undershoots errors present in the data. As explained in Section~\ref{bicycle_data}, the two data sets partly make use of different data models. Moreover, the GeoDanmark data set in its current shape already has been shown unsuitable for routing \citep{septima_geodanmark_2019}. We therefore expect to see some differences in network fragmentation at the local and possibly also at the global level.

\begin{figure}[H]
\includegraphics[width=0.999\textwidth]{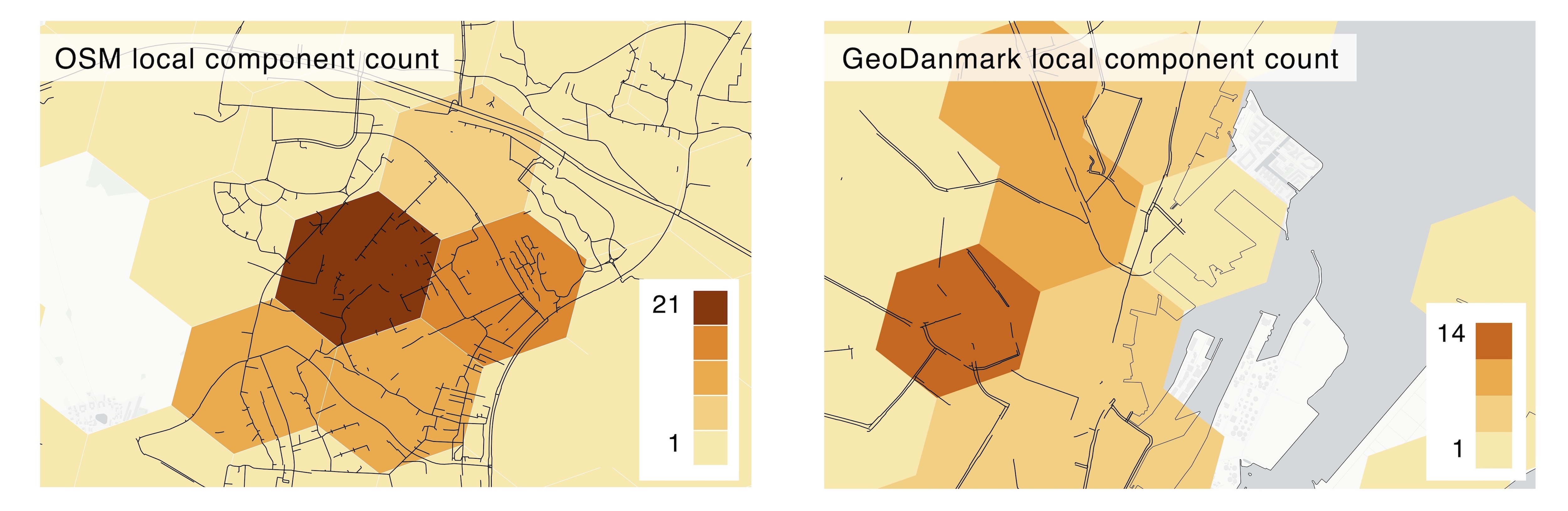}
\caption{\textbf{Local component count}. Examples of areas with a high local component count in OSM data (left) and GeoDanmark data (right). The local component count for both data sets is relatively homogeneous, but with a few outliers with many disconnected components.}
\label{fig:comp_count}
\end{figure}

The ratio of kilometers of bicycle infrastructure to number of network components (Table~\ref{table:extrinsic_table}) is almost identical between OSM and GeoDanmark, with a ratio of \num{1.94} and \num{1.98}, respectively. However, large discrepancies arise in the fragmentation at local (grid cell) level, i.e.~by counting how many different disconnected component a grid cell intersects with. In some locations OSM has a much higher number of disconnected components than GeoDanmark, in other locations the opposite is true. The value range for the local component count is notably wider in OSM (\num{1}-\num{21}) than in GeoDanmark (\num{1}-\num{14}), see Figs.~\ref{fig:comp_count} and S6. 

A high number of disconnected components in close proximity to each other mostly occurs due to very detailed mapping, where e.g.~bicycle lanes of a few meters length result in several disconnected components on the same road. For OSM, erroneous fragmentation also occurs because of missing tags of bicycle infrastructure, which leads to many small missing links and consequently disconnected components. In GeoDanmark, disconnected components often occur due to the data model, which uses separate geometries on each side of a road with no connecting links at e.g.~intersections.

\begin{figure}[H]
\includegraphics[width=0.999\textwidth]{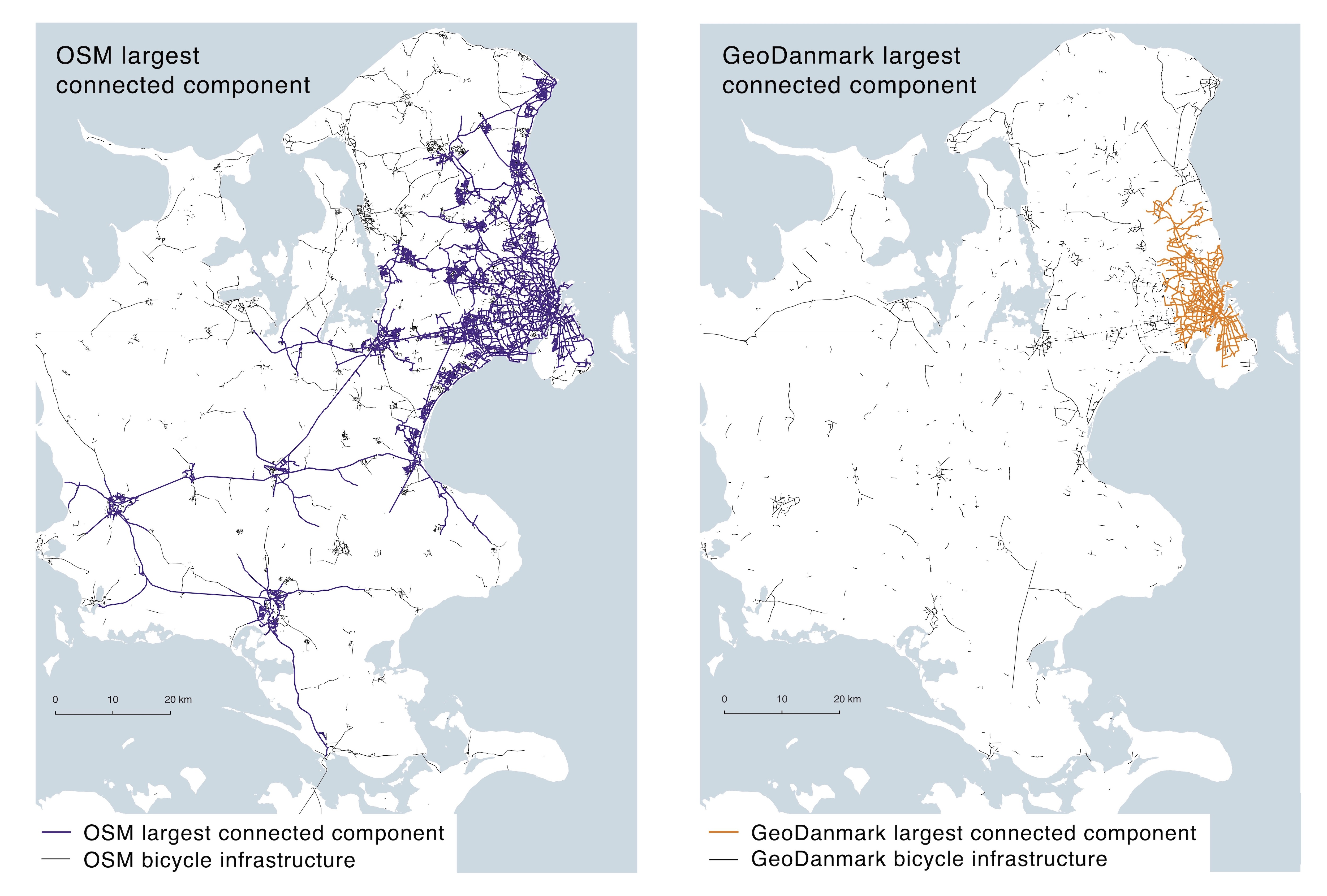}
\caption{\textbf{Largest connected components}. Left: Largest connected component in OSM (purple). Right: Largest connected component in GeoDanmark data (orange)}
\label{fig:lcc}
\end{figure}

Although OSM has a higher maximum count of local disconnected components (Fig.~\ref{fig:comp_count}), its largest connected component is bigger than GeoDanmark's.
The OSM network can therefore be considered more connected. The largest connected components of OSM and GeoDanmark are \num{3433} and \num{1018} \si{km} long and represent \num{22}{\%} and \num{12}{\%} of the total network length, respectively (Table~\ref{table:extrinsic_table} and Fig.~\ref{fig:lcc}). 

\clearpage

\begin{figure}[h]
\centering
\includegraphics[width=0.6\textwidth]{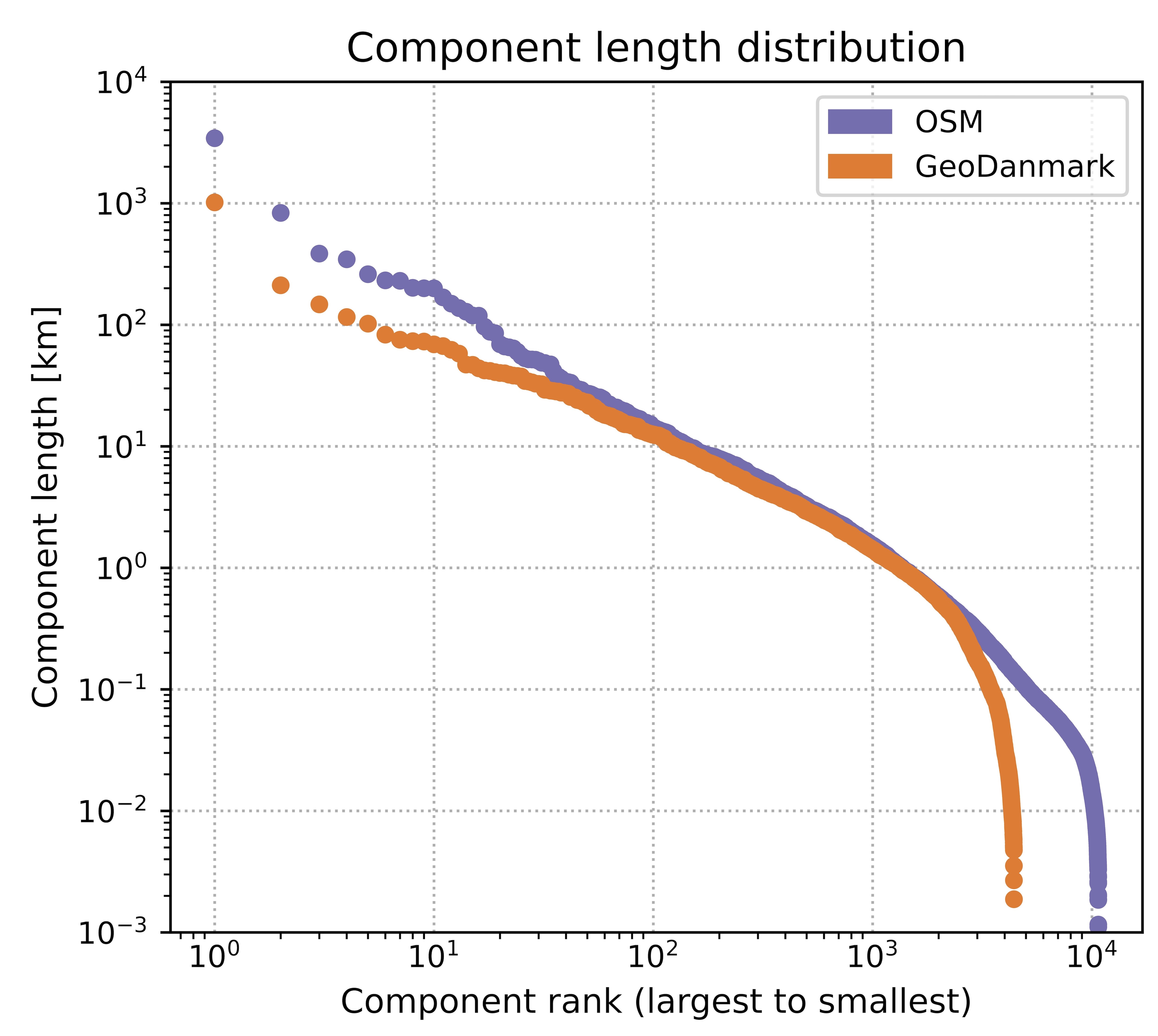}
\caption{Zipf plot ranking the length of OSM components and GeoDanmark components on a log-log scale.}
\label{fig:zipf}
\end{figure}

The differences in network fragmentation can also be seen in the Zipf plot (Fig.~\ref{fig:zipf}), which ranks the lengths of all components by descending order on a log-log scale. For both data sets, the plot shows several data points which represent components that are much larger than the remainder of the components (top left in Fig.~\ref{fig:zipf}). Compared to the GeoDanmark data, the OSM data however have several very large components (starting at the leftmost top marker at rank \si{10^0=1}), and the second highest ranked OSM component is the same size as the highest ranked component in the GeoDanmark data. Although the aggregated values (ratio of kilometers of bicycle infrastructure to number of network components) indicate that OSM and GeoDanmark data have a similar fragmentation, these aggregate values cover a substantial variation in the distribution of component size (Fig.~S6). At grid cell level, the OSM data have more tiny components (the lowest ranked components with length \si{10^{-3} = 0.001} km, see purple tail on the right side in Fig.~\ref{fig:zipf}), but also more very large components with several thousand kilometers of bicycle infrastructure. The OSM network thus both contains large components which can support a cycling network analysis, but also many components which are too small for any meaningful analysis of e.g.~cycling accessibility. In summary, disconnected components occur in both OSM and GeoDanmark data partly because of the scattered nature of the actual infrastructure, but also due to missing links and imprecisely mapped geometries. 

\begin{figure}[H]
\centering
\includegraphics[width=0.999\textwidth]{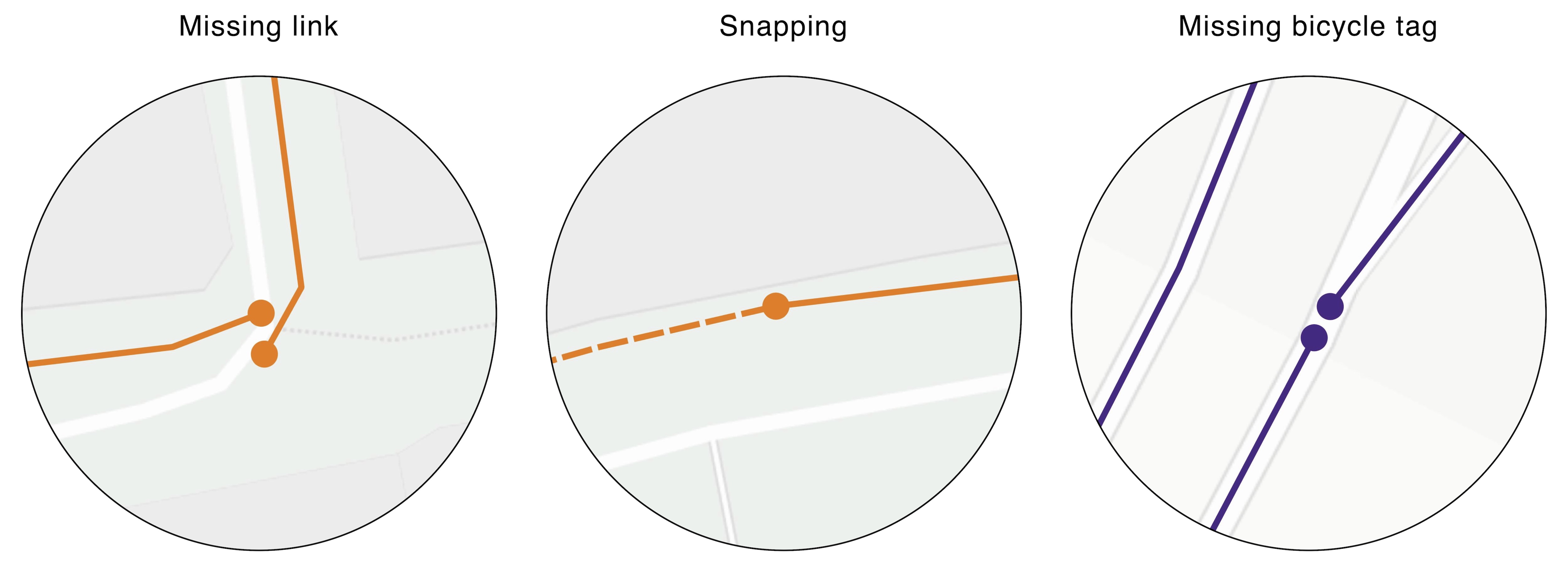}
\caption{\textbf{Different causes for undershoots}. Undershoots detected in GeoDanmark data due to unconnected infrastructure (left). Undershoots because of snapping issues in GeoDanmark data (middle): due to linestrings not being properly connected, the solid and dashed line are not actually connected. Undershoots detected in OSM data (right) due to a missing bicycle tag in underlying road network.}
\label{fig:detected_undershoots}
\end{figure}

As a final aspect of the network structure analysis, we compare the number and locations of undershoots in the two data sets. GeoDanmark data contains \num{339} undershoots, which is much more compared to OSM's \num{157} undershoots, particularly when considering that the GeoDanmark network is only half as long as the OSM network. Based on manual verification of a randomly drawn sample of undershoots in the OSM and GeoDanmark data, we find that a threshold of three meters leads to correct identification of most undershoots with only few false positives: out of \num{40} inspected OSM undershoots, \num{36} were correctly classified as undershoots, \num{4} were false positives. For the inspected GeoDanmark undershoots, \num{26} out of \num{40} were correctly classified, \num{13} were undershoots introduced by the GeoDanmark data model, and \num{1} was a false positive. Through the manual inspection of the undershoots subset, we also find that undershoots appear in OSM and GeoDanmark for different reasons. In OSM, most undershoots are due to missing tags, where small segments of the road network have not been tagged as having dedicated bicycle infrastructure (Fig.~\ref{fig:detected_undershoots}). Due to the differing data model used in GeoDanmark, undershoots mostly appear there because of geometry errors, such as snapping issues and missing links (Fig.~\ref{fig:detected_undershoots}). The undershoots are somewhat unevenly distributed across Denmark within each data set (Fig.~S7), but do not exhibit any significant spatial clustering. The discrepancy in the locations of undershoots between the two data sets suggest that the undershoot indeed are errors, rather than the consequence of a precise mapping of a fragmented infrastructure network. 

In conclusion, both data sets suffer from network fragmentation and topological errors, which poses a problem for network-oriented applications, such as bicycle routing. Due to the diverse reasons for these errors, we cannot issue any universal recommendation for achieving high-quality, routable bicycle infrastructure data. Nevertheless, our findings underline the importance of detailed data quality assessments. Lastly, judging by the sizes of the largest connected components, the OSM network is more connected, and thus \emph{more} suitable for e.g.~routing and accessibility analysis than the GeoDanmark data set. 

\subsection{Results for OSM tags}
\begin{figure}[H]
\centering
\includegraphics[width=0.999\textwidth]{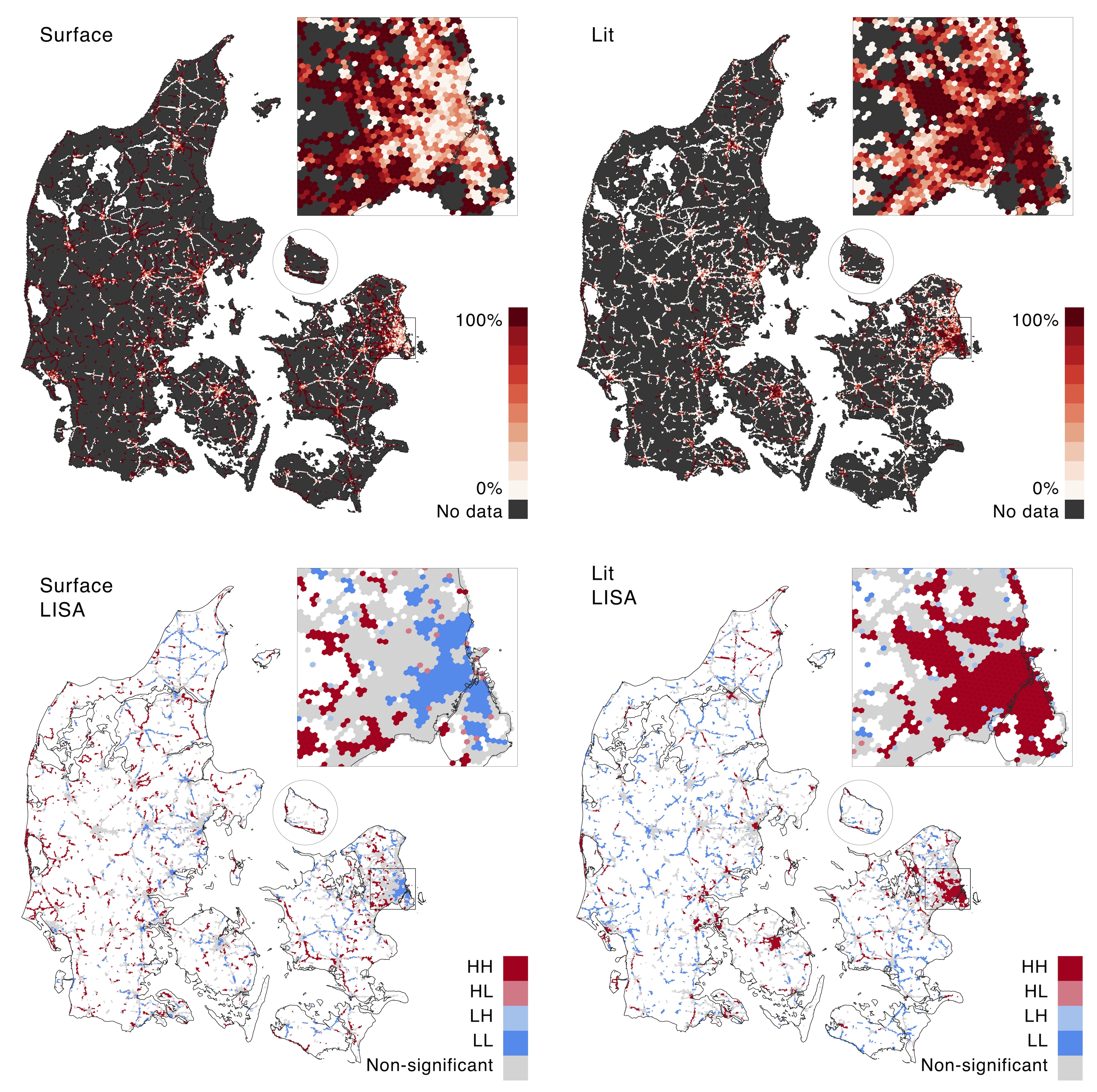}
\caption{\textbf{OSM tag completeness}. Differences in OSM tag completeness: Percent length of network geometries with information for the tag `surface' (top left) and `lit' (top right). In some areas, there is a negative relationship between the tag completeness for different tags. Bottom row: Statistically significant (p<\num{0.05}) clusters of share of infrastructure with the `surface' tag (bottom left) and `lit' tag (bottom right). In several locations, areas with a statistically significant cluster of low `surface' tagging coincide with clusters of high rates of `lit' tagging.}
\label{fig:osm_tags_sa}
\end{figure}

The completeness of OSM tags related to cycling follows notable spatial patterns across Denmark. For all included tags (\textit{`surface'/`cycleway:surface'}; \textit{`width'/`cycleway:width'}; \textit{`lit'}; \textit{`maxspeed'}) we find a large spatial heterogeneity, exhausting the full range from \num{0} - \num{100}\si{\%} local tag completeness (Fig.~S8). Notably, the distribution of tag completeness for the different tags is not random, but instead shows clear clusters of low and high tag completeness (Fig.~\ref{fig:osm_tags_sa}). 

A test for spatial autocorrelation, using the same k-nearest neighbors (k = 6) spatial weight matrix as in Section~\ref{methods_evaluating}, is positive for all tags (Fig.~S9). Interestingly, although the completeness of all the investigated tags has evident clustering tendencies, the locations of clusters with low or high tag completeness follow very dissimilar patterns for different tags, with some tags showing almost reverse patterns. For example, for the bicycle infrastructure within the city of Copenhagen, values for the `surface'/`cycleway:surface' tags are mostly missing, while the `lit' tag (for street light) mostly is present (Fig.~\ref{fig:osm_tags_sa}). Based on the clusters from the spatial autocorrelation (Fig.~\ref{fig:osm_tags_sa}), this pattern is also visible in several other larger towns across the country, with a cluster of low use of the `surface' tag in the town centers coinciding with a cluster of high use of the `lit' tag. More specifically, out of the \num{2460} hex grid cells determined to be in a hot-spot for the `lit' tag, \num{37}{\%} are also in a `surface' cold-spot. On the other hand, \num{38}{\%} of the \num{2396} hex grid cells in a `surface' cold-spot are also in a `lit' hot-spot.

Missing tags are a hindrance for detailed mappings of bicycle conditions \citep{wasserman_evaluating_2019}. Our findings, however, also show that if the presence of tags is to be used as a quality indicator, the type of tag has to be chosen with great care, since different tags have very different levels of completeness. Further, our results reveal that missing tags should not be interpreted as a lack of mapping efforts. Instead, the absence of a specific tag might indicate that it was not deemed relevant by the contributors. For example, the lack of information about the surface of bicycle infrastructure in some city centers can be explained by the fact that dedicated bicycle infrastructure in Danish cities almost always has a paved surface, usually asphalt. For bicycle infrastructure mapped as an attribute to the road center line, the surface of a `cycleway' is therefore often assumed to be the same as of the main road. Tagging completeness is moreover not necessarily an indicator of a lack of mapping efforts, since a high number of contributors editing a OSM feature is no guarantee for many tags being added \citep{mooney_annotation_2012}.

\clearpage

\clearpage

\section{Discussion} \label{discussion}

In this section, we leverage the results from the Denmark case study to issue several recommendations for data quality improvements. We then discuss the limitations of this study and our suggestions for future work.

\subsection{Recommendations for data quality improvements}
An analysis of spatial data quality has two main applications: identifying and rectifying errors and informing data collection efforts. An initial requirement for an analysis like the one presented here is open data. In the case of Denmark, open data on bicycle infrastructure is provided by OSM and GeoDanmark, but up-to-date open data is often unavailable \citep{nelson_crowdsourced_2021}. Given available open data, we provide recommendations for both immediate improvements to the two data sets and more long-term upgrades of data collection and processing:

\textbf{Data conflation.} Our findings suggest that outside of the main urban centers in Denmark, a conflation of OSM and GeoDanmark data (i.e.~merging the two data sets) is necessary to achieve a (more) complete data set. For both OSM and GeoDanmark data it might furthermore be beneficial to close network gaps below a certain distance threshold, or to convert the network to a coarser scale to circumvent inconsistencies from topological errors and smaller missing links, as seen in e.g.~\citet{schoner_missing_2014} and \citet{reggiani_understanding_2021}.

\textbf{Strategic mapping efforts.} Our results indicate that there is a need for more strategic data collection and mapping efforts in Denmark, particularly in areas with large discrepancies between OSM and GeoDanmark data. For OSM, unprotected bicycle infrastructure appears to be particularly under-mapped outside of urban centers. 
According to a study by \citet{hvingel_gode_2023}, missing GeoDanmark bicycle infrastructure data can mostly be addressed by using a more precise labelling that identifies all paths dedicated to cycling.

\textbf{Consistent standards and classifications.} Our comparison of two data sets using a combination of different mapping approaches and various infrastructure classifications also highlights the need for more consistent standards and classifications in bicycle infrastructure data. For GeoDanmark data, we recommend an improved classification of bicycle infrastructure that indicates both the protection level and whether the bicycle infrastructure is running in parallel with a road with motorized traffic. For OSM, an explicit tagging of whether bicycle infrastructure is part of a road with motorized traffic would likewise be an improvement. In theory, bicycle infrastructure that is not part of a road should be marked with the tag `highway = cycleway' \citep{openstreetmap_taghighwaycycleway_2023}, but in practice this tag is used both for separate protected bicycle tracks and tracks that run in parallel with a road.

\textbf{Topological consistency.} Lastly, there is a need for more automated enforcement of topological consistency. For OSM, this could for example include an automated detection of short stretches with no bicycle tag on a road that otherwise has been tagged as having bicycle infrastructure. For GeoDanmark, automatic snapping of geometries and automated detecting of missing links between bicycle infrastructure would greatly improve the consistency of the data. Nevertheless, it will require a larger update to the GeoDanmark data model if the data set is to be used for routing, such as adding consistent connections across intersections \citep{septima_geodanmark_2019}.

\subsection{Limitations}
Although we designed our study with the aim of capturing data quality as precisely as possible, there are a few limitations to the precision of our results. The first one originates from the queries used to obtain the subsets of the road networks with dedicated bicycle infrastructure. As explained in Section~\ref{bicycle_data}, OSM and GeoDanmark use different data models and classifications, and there is therefore no feasible method for obtaining exactly corresponding subsets. While this is important to have in mind when interpreting results on e.g.~data completeness, the lack of consistent classification and mapping practices is also an important lesson in itself and highlights the need for standardized classifications. 

Next, the lack of ground truth and data of a known quality means that findings must be interpreted with care. Some familiarity with the study area is thus required to correctly distinguish between errors of omission or commission, as well as to establish whether network fragmentation is a consequence of low quality \textit{data} or low quality \textit{infrastructure}. Since our study is limited to available, open data, there might exist more high quality data sets in e.g.~road management systems, which however are not available to researchers, bicycle advocates, routing applications, etc. 

Finally, several of the methods in this study make use of customizable settings, such as the distance threshold for undershoots or the maximum distance and angle between corresponding segments in the feature matching process. The best value for these settings depends on the specific data sets and the local context, but will in any case be a potential source of error, since there rarely will be one single and  unambiguous threshold applicable to the entire data set. Moreover, many of the exact results are aggregated and generalized to the grid cell resolution, which, as all spatial data aggregation, can exaggerate or conceal spatial patterns. The most important outcome of the analysis are thus not the exact differences in e.g.~infrastructure density or the precise number of undershoots, but rather the general patterns in data quality and what they can tell us about the state of bicycle infrastructure data in Denmark.

\subsection{Future research}
More work remains within the area of quality assurance of bicycle infrastructure data. Our study has identified some relevant quality issues in the examined bicycle infrastructure data, but tools for actually improving the quality in an automated or efficient manner are still lacking. Therefore, developing reproducible and automated methods for e.g.~conflating bicycle infrastructure data sets from different sources and data models would be a valuable contribution to the fields of bicycle research and sustainable mobility.

Further, there is still no straightforward method to distinguish between gaps and missing links in the data and in the infrastructure. There are suggestions that using street view imagery and image recognition could help solve this problem \citep{biljecki_street_2021, ding_towards_2021, saxton_mapping_2022}, but these methods are still at an early stage of development. 

Finally, more research on standardizing and homogenizing mapping and classifications of bicycle infrastructure is needed in order to ensure comparability and compatibility across data sets. Several local projects on improving the quality and consistency of bicycle infrastructure data classifications are already ongoing in, for example, Canada and Denmark \citep{winters_canadian_2022, hvingel_gode_trafik_2023}.

\section{Conclusion}
\label{conclusion}
In this study, we have examined the spatial data quality of OSM and GeoDanmark, two different data sets on bicycle infrastructure in Denmark, to understand whether these data sets can support a network-based analysis of cycling conditions. Our results reveal large and heterogeneous spatial variations in data completeness and consistency, meaning that the fitness for purpose of the data depends on the geographical location. The highest data quality in terms of data completeness and concordance between the two data sets is found in cities and larger towns. For more rural areas, we found that the information in the two data sets is not sufficient to confidently detect neither how much bicycle infrastructure exists, nor exactly where it is located. For the use case of a network-based analysis of Danish bicycle infrastructure, a conflation of the two data sets is a potential solution -- assuming that the differences in data completeness between the two data sets are due to errors of omission, rather than commission.

In their network structure, both data sets display some unwanted fragmentation due to missing tags, snapping errors, or missing links. If data are to be used for purposes where network structure and topology are of importance, both OSM and GeoDanmark data will thus require some preprocessing, such as closing gaps at intersections or transforming the network to a coarser resolution. The OSM data not only contain more bicycle infrastructure than the GeoDanmark data, but also seem to be better suited for routing and connectivity applications, based on the larger connected components and the smaller number of topology errors. Due to the data model used in GeoDanmark, with all infrastructure mapped with separate geometries rather than a road center line mapping, the GeoDanmark data however offer a higher positional accuracy. This is useful for use cases where the exact location of infrastructure is relevant. 

Due to the lack of ground truth data, no claims about the exact spatial data quality can be made, but we can conclude that open data on dedicated bicycle infrastructure in Denmark still is of an insufficient quality for most use cases, due to the uncertainty of both the extent and location of bicycle infrastructure. We have furthermore demonstrated that the commonly used method of comparing data completeness based on density differences can obscure substantial differences in the exact infrastructure mapped in different data sets. Our study reveals that for more exact and detailed measurements of differences in data completeness, it is necessary to apply more computationally intensive methods like feature matching.

In order to overcome the challenges posed by insufficient data quality, research on networks of dedicated bicycle infrastructure has to either include a substantial amount of automated and/or manual data verification, or to perform analysis of bicycle conditions at a coarser spatial scale, which leads to more uncertainties than desired. However, the results and conclusions presented in this paper are valid for data sets with only dedicated bicycle infrastructure; when considering the entire road network, both data sets are substantially less fragmented.

Data quality improvement is not an end goal in itself, but only a means to an end.  Working towards improving the quality of the available data on bicycle conditions means supporting more data-informed bicycle research and planning. Efforts to improve conditions for active mobility should not be hindered by a lack of reliable data on foundational aspects such as the extent and location of existing infrastructure.

\section*{Acknowledgements}
Data from © OpenStreetMap contributors © GeoDanmark © SDFI © Statistics Denmark. © European Commission. Thanks to all OSM contributors for helping make spatial data open and free. We acknowledge support by the Danish Ministry of Transport.
 
\section*{Conflicts of interest}
The authors declare no conflict of interest.

\section*{Reproducibility and data availability}

All code used in the analysis can be found at ~\url{https://github.com/anerv/BikeDNA_BIG} and ~\url{https://github.com/anerv/bikedna_dk_analysis}. See \url{https://github.com/anerv/BikeDNA} for the original version of \texttt{BikeDNA}. All input data and results from the analysis can be found at \url{https://doi.org/10.5281/zenodo.8340383} and \url{https://doi.org/10.5281/zenodo.10185500}.

\bibliography{main.bib}

\begin{thebibliography}{87}
\providecommand{\natexlab}[1]{#1}
\providecommand{\url}[1]{\texttt{#1}}
\expandafter\ifx\csname urlstyle\endcsname\relax
  \providecommand{\doi}[1]{doi: #1}\else
  \providecommand{\doi}{doi: \begingroup \urlstyle{rm}\Url}\fi

\bibitem[Almendros-Jiménez and Becerra-Terón(2018)]{almendros-jimenez_analyzing_2018}
Jesús~M. Almendros-Jiménez and Antonio Becerra-Terón.
\newblock Analyzing the {Tagging} {Quality} of the {Spanish} {OpenStreetMap}.
\newblock \emph{ISPRS International Journal of Geo-Information}, 7\penalty0 (8):\penalty0 323, August 2018.
\newblock ISSN 2220-9964.
\newblock \doi{10.3390/ijgi7080323}.
\newblock URL \url{https://www.mdpi.com/2220-9964/7/8/323}.
\newblock Number: 8 Publisher: Multidisciplinary Digital Publishing Institute.

\bibitem[Anselin(1995)]{anselin_local_1995}
Luc Anselin.
\newblock Local {Indicators} of {Spatial} {Association}—{LISA}.
\newblock \emph{Geographical Analysis}, 27\penalty0 (2):\penalty0 93--115, 1995.
\newblock ISSN 1538-4632.
\newblock \doi{10.1111/j.1538-4632.1995.tb00338.x}.
\newblock URL \url{https://onlinelibrary.wiley.com/doi/abs/10.1111/j.1538-4632.1995.tb00338.x}.
\newblock \_eprint: https://onlinelibrary.wiley.com/doi/pdf/10.1111/j.1538-4632.1995.tb00338.x.

\bibitem[Barrington-Leigh and Millard-Ball(2017)]{barrington-leigh_worlds_2017}
Christopher Barrington-Leigh and Adam Millard-Ball.
\newblock The world's user-generated road map is more than 80\% complete.
\newblock \emph{PloS One}, 12\penalty0 (8):\penalty0 e0180698, 2017.
\newblock ISSN 1932-6203.
\newblock \doi{10.1371/journal.pone.0180698}.

\bibitem[Barron et~al.(2014)Barron, Neis, and Zipf]{barron_comprehensive_2014}
Christopher Barron, Pascal Neis, and Alexander Zipf.
\newblock A {Comprehensive} {Framework} for {Intrinsic} {OpenStreetMap} {Quality} {Analysis}: {A} {Comprehensive} {Framework} for {Intrinsic} {OpenStreetMap} {Quality} {Analysis}.
\newblock \emph{Transactions in GIS}, 18\penalty0 (6):\penalty0 877--895, December 2014.
\newblock ISSN 13611682.
\newblock \doi{10.1111/tgis.12073}.
\newblock URL \url{https://onlinelibrary.wiley.com/doi/10.1111/tgis.12073}.

\bibitem[Biljecki and Ito(2021)]{biljecki_street_2021}
Filip Biljecki and Koichi Ito.
\newblock Street view imagery in urban analytics and {GIS}: {A} review.
\newblock \emph{Landscape and Urban Planning}, 215:\penalty0 104217, November 2021.
\newblock ISSN 0169-2046.
\newblock \doi{10.1016/j.landurbplan.2021.104217}.
\newblock URL \url{https://www.sciencedirect.com/science/article/pii/S0169204621001808}.

\bibitem[Biljecki et~al.(2023)Biljecki, Chow, and Lee]{biljecki_quality_2023}
Filip Biljecki, Yoong~Shin Chow, and Kay Lee.
\newblock Quality of crowdsourced geospatial building information: {A} global assessment of {OpenStreetMap} attributes.
\newblock \emph{Building and Environment}, 237:\penalty0 110295, June 2023.
\newblock ISSN 0360-1323.
\newblock \doi{10.1016/j.buildenv.2023.110295}.
\newblock URL \url{https://www.sciencedirect.com/science/article/pii/S0360132323003220}.

\bibitem[Boeing(2017)]{boeing_osmnx_2017}
Geoff Boeing.
\newblock {OSMnx}: {New} methods for acquiring, constructing, analyzing, and visualizing complex street networks.
\newblock \emph{Computers, Environment and Urban Systems}, 65:\penalty0 126--139, September 2017.
\newblock ISSN 01989715.
\newblock \doi{10.1016/j.compenvurbsys.2017.05.004}.
\newblock URL \url{https://linkinghub.elsevier.com/retrieve/pii/S0198971516303970}.

\bibitem[Brando and Bucher(2010)]{brando_quality_2010}
Carmen Brando and Bénédicte Bucher.
\newblock Quality in {User} {Generated} {Spatial} {Content}: {A} {Matter} of {Specifications}.
\newblock In \emph{13th {AGILE} {International} {Conference} on {Geographic} {Information} {Science}}, Guimaraes, Portugal, April 2010.
\newblock URL \url{https://hal.archives-ouvertes.fr/hal-02435222}.

\bibitem[Brovelli et~al.(2017)Brovelli, Minghini, Molinari, and Mooney]{brovelli_towards_2017}
Maria~Antonia Brovelli, Marco Minghini, Monia Molinari, and Peter Mooney.
\newblock Towards an {Automated} {Comparison} of {OpenStreetMap} with {Authoritative} {Road} {Datasets}.
\newblock \emph{Transactions in GIS}, 21\penalty0 (2):\penalty0 191--206, April 2017.
\newblock ISSN 1361-1682, 1467-9671.
\newblock \doi{10.1111/tgis.12182}.
\newblock URL \url{https://onlinelibrary.wiley.com/doi/10.1111/tgis.12182}.

\bibitem[Buehler and Dill(2016)]{buehler_bikeway_2016}
Ralph Buehler and Jennifer Dill.
\newblock Bikeway {Networks}: {A} {Review} of {Effects} on {Cycling}.
\newblock \emph{Transport Reviews}, 36\penalty0 (1):\penalty0 9--27, January 2016.
\newblock ISSN 0144-1647.
\newblock \doi{10.1080/01441647.2015.1069908}.
\newblock URL \url{https://doi.org/10.1080/01441647.2015.1069908}.
\newblock Publisher: Routledge \_eprint: https://doi.org/10.1080/01441647.2015.1069908.

\bibitem[Carlino et~al.(2023)Carlino, Li, and Kirk]{carlino_b_2023}
Dustin Carlino, Yuwen Li, and Michael Kirk.
\newblock A/{B} {Street}, February 2023.
\newblock URL \url{https://github.com/a-b-street/abstreet}.
\newblock original-date: 2018-06-04T00:44:43Z.

\bibitem[CHIPS(2019)]{chips_european_2019}
CHIPS.
\newblock European map for {Potential} {Cycle} {Highways}, 2019.
\newblock URL \url{https://cyclehighways.eu/index.php?id=129}.

\bibitem[CycleStreets(2023)]{cyclestreets_cyclestreets_2023}
CycleStreets.
\newblock {CycleStreets} - {UK}-wide cycle routing and intelligence, 2023.
\newblock URL \url{https://m.cyclestreets.net/#5/53.78/-2.37}.

\bibitem[{Datafordeler}(2023)]{datafordeler_datafordelerdk_2023}
{Datafordeler}.
\newblock Datafordeler.dk, 2023.
\newblock URL \url{https://datafordeler.dk/}.

\bibitem[Degrossi et~al.(2018)Degrossi, Porto~de Albuquerque, Santos~Rocha, and Zipf]{degrossi_taxonomy_2018}
Lívia~Castro Degrossi, João Porto~de Albuquerque, Roberto~dos Santos~Rocha, and Alexander Zipf.
\newblock A taxonomy of quality assessment methods for volunteered and crowdsourced geographic information.
\newblock \emph{Transactions in GIS}, 22\penalty0 (2):\penalty0 542--560, 2018.
\newblock ISSN 1467-9671.
\newblock \doi{10.1111/tgis.12329}.
\newblock URL \url{https://onlinelibrary.wiley.com/doi/abs/10.1111/tgis.12329}.

\bibitem[Devillers et~al.(2007)Devillers, Bédard, Jeansoulin, and Moulin]{devillers_towards_2007}
R.~Devillers, Y.~Bédard, R.~Jeansoulin, and B.~Moulin.
\newblock Towards spatial data quality information analysis tools for experts assessing the fitness for use of spatial data.
\newblock \emph{International Journal of Geographical Information Science}, 21\penalty0 (3):\penalty0 261--282, March 2007.
\newblock ISSN 1365-8816.
\newblock \doi{10.1080/13658810600911879}.
\newblock URL \url{https://doi.org/10.1080/13658810600911879}.
\newblock Publisher: Taylor \& Francis \_eprint: https://doi.org/10.1080/13658810600911879.

\bibitem[Ding et~al.(2021)Ding, Fan, and Gong]{ding_towards_2021}
Xuan Ding, Hongchao Fan, and Jianya Gong.
\newblock Towards generating network of bikeways from {Mapillary} data.
\newblock \emph{Computers, Environment and Urban Systems}, 88:\penalty0 101632, July 2021.
\newblock ISSN 0198-9715.
\newblock \doi{10.1016/j.compenvurbsys.2021.101632}.
\newblock URL \url{https://www.sciencedirect.com/science/article/pii/S0198971521000399}.

\bibitem[ECF(2022)]{ecf_integrated_2022}
ECF.
\newblock “{Integrated} {Cycling} {Planning} {Guide}”: {An} {EU} {CYCLE} tool for building regional cycle networks, January 2022.
\newblock URL \url{https://ecf.com/news-and-events/news/integrated-cycling-planner-guide-eu-cycle-tool-building-regional-cycle-networks}.

\bibitem[EEA(2022)]{eea_greenhouse_2022}
EEA.
\newblock Greenhouse gas emissions from transport in {Europe}, 2022.
\newblock URL \url{https://www.eea.europa.eu/ims/greenhouse-gas-emissions-from-transport}.

\bibitem[Eudaly et~al.(2020)Eudaly, Warner, Pearce, Igarta, Geller, Phillips, Serritella, Gastaldi, Valle, and Slyman]{eudaly_portland_2020}
Chloe Eudaly, Chris Warner, Art Pearce, Denver Igarta, Roger Geller, Taylor Phillips, Michael Serritella, Gena Gastaldi, Shane Valle, and Owen Slyman.
\newblock Portland {Bicycle} {Plan} for 2030 - 2019 {Progress} {Report}.
\newblock Technical report, Portland Bureau of Transportation, 2020.

\bibitem[{European Commission}(2021)]{ec_new_2021}
{European Commission}.
\newblock The {New} {EU} {Urban} {Mobility} {Framework}.
\newblock Technical report, European Commission, Brussels, November 2021.
\newblock URL \url{https://transport.ec.europa.eu/system/files/2021-12/com_2021_811_the-new-eu-urban-mobility.pdf}.

\bibitem[{European Commission}(2023)]{european_commission_european_2023}
{European Commission}.
\newblock European {Declaration} on {Cycling}.
\newblock Technical report, European Commission, Brussels, 2023.
\newblock URL \url{https://transport.ec.europa.eu/system/files/2023-10/European_Declaration_on_Cycling.pdf}.

\bibitem[Ferster et~al.(2020)Ferster, Fischer, Manaugh, Nelson, and Winters]{ferster_using_2020}
Colin Ferster, Jaimy Fischer, Kevin Manaugh, Trisalyn Nelson, and Meghan Winters.
\newblock Using {OpenStreetMap} to inventory bicycle infrastructure: {A} comparison with open data from cities.
\newblock \emph{International Journal of Sustainable Transportation}, 14\penalty0 (1):\penalty0 64--73, January 2020.
\newblock ISSN 1556-8318, 1556-8334.
\newblock \doi{10.1080/15568318.2018.1519746}.
\newblock URL \url{https://www.tandfonline.com/doi/full/10.1080/15568318.2018.1519746}.

\bibitem[Fleischmann(2019)]{fleischmann_momepy_2019}
Martin Fleischmann.
\newblock momepy: {Urban} {Morphology} {Measuring} {Toolkit}.
\newblock \emph{Journal of Open Source Software}, 4\penalty0 (43):\penalty0 1807, November 2019.
\newblock ISSN 2475-9066.
\newblock \doi{10.21105/joss.01807}.
\newblock URL \url{https://joss.theoj.org/papers/10.21105/joss.01807}.

\bibitem[Fonte(2017)]{fonte_assessing_2017}
Cidalia Fonte.
\newblock Assessing {VGI} {Data} {Quality}.
\newblock \emph{Mapping and the Citizen Sensor}, pages 137--163, 2017.
\newblock \doi{10.5334/bbf.g}.
\newblock URL \url{https://www.ubiquitypress.com/site/chapters/10.5334/bbf.g/}.

\bibitem[Forghani and Delavar(2014)]{forghani_quality_2014}
Mohammad Forghani and Mahmoud~Reza Delavar.
\newblock A {Quality} {Study} of the {OpenStreetMap} {Dataset} for {Tehran}.
\newblock \emph{ISPRS International Journal of Geo-Information}, 3\penalty0 (2):\penalty0 750--763, June 2014.
\newblock ISSN 2220-9964.
\newblock \doi{10.3390/ijgi3020750}.
\newblock URL \url{https://www.mdpi.com/2220-9964/3/2/750}.
\newblock Number: 2 Publisher: Multidisciplinary Digital Publishing Institute.

\bibitem[Fosgerau et~al.(2023)Fosgerau, Łukawska, Paulsen, and Rasmussen]{fosgerau_bikeability_2023}
Mogens Fosgerau, Mirosława Łukawska, Mads Paulsen, and Thomas~Kjær Rasmussen.
\newblock Bikeability and the induced demand for cycling.
\newblock \emph{Proceedings of the National Academy of Sciences}, 120\penalty0 (16):\penalty0 e2220515120, April 2023.
\newblock \doi{10.1073/pnas.2220515120}.
\newblock URL \url{https://www.pnas.org/doi/10.1073/pnas.2220515120}.
\newblock Publisher: Proceedings of the National Academy of Sciences.

\bibitem[Furth et~al.(2016)Furth, Mekuria, and Nixon]{furth_network_2016}
Peter~G. Furth, Maaza~C. Mekuria, and Hilary Nixon.
\newblock Network {Connectivity} for {Low}-{Stress} {Bicycling}.
\newblock \emph{Transportation Research Record}, 2587\penalty0 (1):\penalty0 41--49, January 2016.
\newblock ISSN 0361-1981.
\newblock \doi{10.3141/2587-06}.
\newblock URL \url{https://doi.org/10.3141/2587-06}.
\newblock Publisher: SAGE Publications Inc.

\bibitem[GeoDanmark(2020)]{geodanmark_produktion_2020}
GeoDanmark.
\newblock Produktion og vedligehold, 2020.
\newblock URL \url{https://www.geodanmark.dk/anvend-geodata/vedligehold-og-produktion/}.

\bibitem[GeoDanmark(2023)]{geodanmark_danmarks_2023}
GeoDanmark.
\newblock Danmarks {Geografi} - {GeoDanmark}, 2023.
\newblock URL \url{https://dataforsyningen.dk/data/3563}.

\bibitem[{Geofabrik}(2020)]{geofabrik_our_2020}
{Geofabrik}.
\newblock Our {Download} {Server}, 2020.
\newblock URL \url{https://www.geofabrik.de/data/download.html}.

\bibitem[Getis(2007)]{getis_reflections_2007}
Arthur Getis.
\newblock Reflections on spatial autocorrelation.
\newblock \emph{Regional Science and Urban Economics}, 37\penalty0 (4):\penalty0 491--496, July 2007.
\newblock ISSN 0166-0462.
\newblock \doi{10.1016/j.regsciurbeco.2007.04.005}.
\newblock URL \url{https://www.sciencedirect.com/science/article/pii/S0166046207000348}.

\bibitem[Graser et~al.(2015)Graser, Straub, and Dragaschnig]{gartner_is_2015}
Anita Graser, Markus Straub, and Melitta Dragaschnig.
\newblock Is {OSM} {Good} {Enough} for {Vehicle} {Routing}? {A} {Study} {Comparing} {Street} {Networks} in {Vienna}.
\newblock In Georg Gartner and Haosheng Huang, editors, \emph{Progress in {Location}-{Based} {Services} 2014}, pages 3--17. Springer International Publishing, Cham, 2015.
\newblock ISBN 978-3-319-11878-9 978-3-319-11879-6.
\newblock \doi{10.1007/978-3-319-11879-6_1}.
\newblock URL \url{http://link.springer.com/10.1007/978-3-319-11879-6_1}.
\newblock Series Title: Lecture Notes in Geoinformation and Cartography.

\bibitem[Gröchenig et~al.(2014)Gröchenig, Brunauer, and Rehrl]{grochenig_estimating_2014}
Simon Gröchenig, Richard Brunauer, and Karl Rehrl.
\newblock Estimating {Completeness} of {VGI} {Datasets} by {Analyzing} {Community} {Activity} {Over} {Time} {Periods}.
\newblock In Joaquín Huerta, Sven Schade, and Carlos Granell, editors, \emph{Connecting a {Digital} {Europe} {Through} {Location} and {Place}}, Lecture {Notes} in {Geoinformation} and {Cartography}, pages 3--18. Springer International Publishing, Cham, 2014.
\newblock ISBN 978-3-319-03611-3.
\newblock \doi{10.1007/978-3-319-03611-3_1}.
\newblock URL \url{https://doi.org/10.1007/978-3-319-03611-3_1}.

\bibitem[Gössling and McRae(2022)]{gossling_subjectively_2022}
Stefan Gössling and Sophia McRae.
\newblock Subjectively safe cycling infrastructure: {New} insights for urban designs.
\newblock \emph{Journal of Transport Geography}, 101:\penalty0 103340, May 2022.
\newblock ISSN 0966-6923.
\newblock \doi{10.1016/j.jtrangeo.2022.103340}.
\newblock URL \url{https://www.sciencedirect.com/science/article/pii/S0966692322000631}.

\bibitem[Haklay(2010)]{haklay_how_good_2010}
Mordechai Haklay.
\newblock How {Good} is {Volunteered} {Geographical} {Information}? {A} {Comparative} {Study} of {OpenStreetMap} and {Ordnance} {Survey} {Datasets}.
\newblock \emph{Environment and Planning B: Planning and Design}, 37\penalty0 (4):\penalty0 682--703, August 2010.
\newblock ISSN 0265-8135, 1472-3417.
\newblock \doi{10.1068/b35097}.
\newblock URL \url{http://journals.sagepub.com/doi/10.1068/b35097}.

\bibitem[Hashemi and Abbaspour(2015)]{hashemi_assessment_2015}
Peyman Hashemi and Rahim~Ali Abbaspour.
\newblock Assessment of {Logical} {Consistency} in {OpenStreetMap} {Based} on the {Spatial} {Similarity} {Concept}.
\newblock In Jamal Jokar~Arsanjani, Alexander Zipf, Peter Mooney, and Marco Helbich, editors, \emph{{OpenStreetMap} in {GIScience}: {Experiences}, {Research}, and {Applications}}, Lecture {Notes} in {Geoinformation} and {Cartography}, pages 19--36. Springer International Publishing, Cham, 2015.
\newblock ISBN 978-3-319-14280-7.
\newblock \doi{10.1007/978-3-319-14280-7_2}.
\newblock URL \url{https://doi.org/10.1007/978-3-319-14280-7_2}.

\bibitem[Hochmair et~al.(2015)Hochmair, Zielstra, and Neis]{hochmair_assessing_2015}
Hartwig~H. Hochmair, Dennis Zielstra, and Pascal Neis.
\newblock Assessing the {Completeness} of {Bicycle} {Trail} and {Lane} {Features} in {OpenStreetMap} for the {United} {States}: {Completeness} of {Bicycle} {Features} in {OpenStreetMap}.
\newblock \emph{Transactions in GIS}, 19\penalty0 (1):\penalty0 63--81, February 2015.
\newblock ISSN 13611682.
\newblock \doi{10.1111/tgis.12081}.
\newblock URL \url{https://onlinelibrary.wiley.com/doi/10.1111/tgis.12081}.

\bibitem[Hvingel and Jensen(2023{\natexlab{a}})]{hvingel_gode_2023}
Line Hvingel and Thomas Jensen.
\newblock Gode cykeldata til alle.
\newblock \emph{Trafik og veje}, 2023{\natexlab{a}}.
\newblock URL \url{https://www.kl.dk/media/53604/artikel_gode_cykeldata_trafik_og_veje_jan_2023_layout.pdf}.

\bibitem[Hvingel and Jensen(2023{\natexlab{b}})]{hvingel_gode_trafik_2023}
Line Hvingel and Thomas Jensen.
\newblock Gode cykeldata til alle.
\newblock \emph{Teknik \& Miljø}, 123\penalty0 (4):\penalty0 36--39, 2023{\natexlab{b}}.
\newblock URL \url{https://www.kl.dk/media/sbphaxrt/gode_cykeldata_til_alle_teknikogmiljoe04_20230404.pdf}.

\bibitem[Jaramillo et~al.(2022)Jaramillo, Kahn~Ribeiro, Newman, Dhar, Diemuodeke, Kajino, Lee, Nugroho, Ou, Hammer~Strømman, and Whitehead]{jaramillo_transport_2022}
Paula Jaramillo, Suzana Kahn~Ribeiro, Peter Newman, Subash Dhar, Ogheneruona~E Diemuodeke, Tsutomu Kajino, David~Simon Lee, Sudarmanto~Budi Nugroho, Xunmin Ou, Anders Hammer~Strømman, and Jake Whitehead.
\newblock Transport.
\newblock In \emph{Climate {Change} 2022: {Mitigation} of {Climate} {Change}}, pages 1049--1160. Intergovernmental Panel on Climate Change (IPCC), 2022.
\newblock URL \url{doi:10.1017/9781009157926.012}.

\bibitem[Kamel and Sayed(2021)]{kamel_impact_2021}
Mohamed~Bayoumi Kamel and Tarek Sayed.
\newblock The impact of bike network indicators on bike kilometers traveled and bike safety: {A} network theory approach.
\newblock \emph{Environment and Planning B: Urban Analytics and City Science}, 48\penalty0 (7):\penalty0 2055--2072, September 2021.
\newblock ISSN 2399-8083.
\newblock \doi{10.1177/2399808320964469}.
\newblock URL \url{https://doi.org/10.1177/2399808320964469}.
\newblock Publisher: SAGE Publications Ltd STM.

\bibitem[Keßler et~al.(2011)Keßler, Trame, and Kauppinen]{kesler_tracking_2011}
Carsten Keßler, Johannes Trame, and Tomi Kauppinen.
\newblock Tracking {Editing} {Processes} in {Volunteered} {Geographic} {Information}: {The} {Case} of {OpenStreetMap}.
\newblock \emph{Proceedings of the Conference on Spatial Information Theory, Workshop: Identifying Objects, Processes and Events in Spatio-Temporally Distributed Data}, page~7, 2011.

\bibitem[Koukoletsos et~al.(2011)Koukoletsos, Haklay, and Ellul]{koukoletsos_automated_2011}
Thomas Koukoletsos, Mordechai~(muki Haklay, and Claire Ellul.
\newblock An automated method to assess {Data} {Completeness} and {Positional} {Accuracy} of {OpenStreetMap}, 2011.
\newblock URL \url{http://www.geog.leeds.ac.uk/groups/geocomp/2011/papers/koukoletsos.pdf}.

\bibitem[Koukoletsos et~al.(2012)Koukoletsos, Haklay, and Ellul]{koukoletsos_assessing_2012}
Thomas Koukoletsos, Mordechai Haklay, and Claire Ellul.
\newblock Assessing {Data} {Completeness} of {VGI} through an {Automated} {Matching} {Procedure} for {Linear} {Data}.
\newblock \emph{Transactions in GIS}, 16\penalty0 (4):\penalty0 477--498, 2012.
\newblock ISSN 1467-9671.
\newblock \doi{10.1111/j.1467-9671.2012.01304.x}.
\newblock URL \url{https://onlinelibrary.wiley.com/doi/abs/10.1111/j.1467-9671.2012.01304.x}.

\bibitem[Lee and Sener(2020)]{lee_emerging_2020}
Kyuhyun Lee and Ipek~N. Sener.
\newblock Emerging data for pedestrian and bicycle monitoring: {Sources} and applications.
\newblock \emph{Transportation Research Interdisciplinary Perspectives}, 4:\penalty0 100095, March 2020.
\newblock ISSN 2590-1982.
\newblock \doi{10.1016/j.trip.2020.100095}.
\newblock URL \url{https://www.sciencedirect.com/science/article/pii/S2590198220300063}.

\bibitem[Lovelace et~al.(2017)Lovelace, Goodman, Aldred, Berkoff, Abbas, and Woodcock]{lovelace_propensity_2017}
Robin Lovelace, Anna Goodman, Rachel Aldred, Nikolai Berkoff, Ali Abbas, and James Woodcock.
\newblock The {Propensity} to {Cycle} {Tool}: {An} open source online system for sustainable transport planning.
\newblock \emph{Journal of Transport and Land Use}, 10\penalty0 (1), January 2017.
\newblock ISSN 1938-7849.
\newblock \doi{10.5198/jtlu.2016.862}.
\newblock URL \url{https://www.jtlu.org/index.php/jtlu/article/view/862}.
\newblock Number: 1.

\bibitem[Lowry and Loh(2017)]{lowry_quantifying_2017}
Michael Lowry and Tracy~Hadden Loh.
\newblock Quantifying bicycle network connectivity.
\newblock \emph{Preventive Medicine}, 95:\penalty0 S134--S140, February 2017.
\newblock ISSN 0091-7435.
\newblock \doi{10.1016/j.ypmed.2016.12.007}.
\newblock URL \url{https://www.sciencedirect.com/science/article/pii/S0091743516304029}.

\bibitem[Mattioli(2021)]{mattioli_chapter_2021}
Giulio Mattioli.
\newblock Chapter {Four} - {Transport} poverty and car dependence: {A} {European} perspective.
\newblock In Rafael H.~M. Pereira and Geneviève Boisjoly, editors, \emph{Advances in {Transport} {Policy} and {Planning}}, volume~8 of \emph{Social {Issues} in {Transport} {Planning}}, pages 101--133. Academic Press, January 2021.
\newblock \doi{10.1016/bs.atpp.2021.06.004}.
\newblock URL \url{https://www.sciencedirect.com/science/article/pii/S2543000921000263}.

\bibitem[Medeiros and Holanda(2019)]{medeiros_solutions_2019}
Gabriel Medeiros and Maristela Holanda.
\newblock Solutions for {Data} {Quality} in {GIS} and {VGI}: {A} {Systematic} {Literature} {Review}.
\newblock In Alvaro Rocha, Hojjat Adeli, Luís~Paulo Reis, and Sandra Costanzo, editors, \emph{New {Knowledge} in {Information} {Systems} and {Technologies}}, Advances in {Intelligent} {Systems} and {Computing}, pages 645--654, Cham, 2019. Springer International Publishing.
\newblock ISBN 978-3-030-16181-1.
\newblock \doi{10.1007/978-3-030-16181-1_61}.

\bibitem[Mekuria et~al.(2012)Mekuria, Furth, and Nixon]{mekuria_low-stress_2012}
M.~C. Mekuria, Peter~G. Furth, and H.~Nixon.
\newblock Low-{Stress} {Bicycling} and {Network} {Connectivity}.
\newblock Technical Report 11-19, Mineta Transportation Institute, 2012.
\newblock URL \url{https://www.semanticscholar.org/paper/Low-Stress-Bicycling-and-Network-Connectivity-Mekuria-Furth/a50063c06112d3eb6aa752dfd362e1bdbc7f1c7e}.

\bibitem[Mennis(2019)]{mennis_problems_2019}
Jeremy Mennis.
\newblock Problems of {Scale} and {Zoning}.
\newblock \emph{Geographic Information Science \& Technology Body of Knowledge}, 2019\penalty0 (Q1), January 2019.
\newblock ISSN 25772848.
\newblock \doi{10.22224/gistbok/2019.1.2}.
\newblock URL \url{https://gistbok.ucgis.org/bok-topics/problems-scale-and-zoning}.

\bibitem[Mooney and Corcoran(2012)]{mooney_annotation_2012}
Peter Mooney and Padraig Corcoran.
\newblock The {Annotation} {Process} in {OpenStreetMap}.
\newblock \emph{Transactions in GIS}, 16\penalty0 (4):\penalty0 561--579, 2012.
\newblock ISSN 1467-9671.
\newblock \doi{10.1111/j.1467-9671.2012.01306.x}.
\newblock URL \url{https://onlinelibrary.wiley.com/doi/abs/10.1111/j.1467-9671.2012.01306.x}.
\newblock \_eprint: https://onlinelibrary.wiley.com/doi/pdf/10.1111/j.1467-9671.2012.01306.x.

\bibitem[Natera~Orozco et~al.(2020)Natera~Orozco, Battiston, Iñiguez, and Szell]{natera_orozco_data-driven_2020}
Luis~Guillermo Natera~Orozco, Federico Battiston, Gerardo Iñiguez, and Michael Szell.
\newblock Data-driven strategies for optimal bicycle network growth.
\newblock \emph{Royal Society Open Science}, 7\penalty0 (12):\penalty0 201130, 2020.
\newblock \doi{10.1098/rsos.201130}.
\newblock URL \url{https://royalsocietypublishing.org/doi/full/10.1098/rsos.201130}.
\newblock Publisher: Royal Society.

\bibitem[Neis et~al.(2012)Neis, Zielstra, and Zipf]{neis_street_2012}
Pascal Neis, Dennis Zielstra, and Alexander Zipf.
\newblock The {Street} {Network} {Evolution} of {Crowdsourced} {Maps}: {OpenStreetMap} in {Germany} 2007–2011.
\newblock \emph{Future Internet}, 4\penalty0 (1):\penalty0 1--21, March 2012.
\newblock ISSN 1999-5903.
\newblock \doi{10.3390/fi4010001}.
\newblock URL \url{https://www.mdpi.com/1999-5903/4/1/1}.
\newblock Number: 1 Publisher: Molecular Diversity Preservation International.

\bibitem[Neis et~al.(2013)Neis, Zielstra, and Zipf]{neis_comparison_2013}
Pascal Neis, Dennis Zielstra, and Alexander Zipf.
\newblock Comparison of {Volunteered} {Geographic} {Information} {Data} {Contributions} and {Community} {Development} for {Selected} {World} {Regions}.
\newblock \emph{Future Internet}, 5\penalty0 (2):\penalty0 282--300, June 2013.
\newblock ISSN 1999-5903.
\newblock \doi{10.3390/fi5020282}.
\newblock URL \url{https://www.mdpi.com/1999-5903/5/2/282}.
\newblock Number: 2 Publisher: Multidisciplinary Digital Publishing Institute.

\bibitem[Nelson et~al.(2021)Nelson, Ferster, Laberee, Fuller, and Winters]{nelson_crowdsourced_2021}
Trisalyn Nelson, Colin Ferster, Karen Laberee, Daniel Fuller, and Meghan Winters.
\newblock Crowdsourced data for bicycling research and practice.
\newblock \emph{Transport Reviews}, 41\penalty0 (1):\penalty0 97--114, January 2021.
\newblock ISSN 0144-1647.
\newblock \doi{10.1080/01441647.2020.1806943}.
\newblock URL \url{https://doi.org/10.1080/01441647.2020.1806943}.

\bibitem[Olmos et~al.(2020)Olmos, Tadeo, Vlachogiannis, Alhasoun, Espinet~Alegre, Ochoa, Targa, and González]{olmos_data_2020}
Luis~E. Olmos, Maria~Sol Tadeo, Dimitris Vlachogiannis, Fahad Alhasoun, Xavier Espinet~Alegre, Catalina Ochoa, Felipe Targa, and Marta~C. González.
\newblock A data science framework for planning the growth of bicycle infrastructures.
\newblock \emph{Transportation Research Part C: Emerging Technologies}, 115:\penalty0 102640, June 2020.
\newblock ISSN 0968-090X.
\newblock \doi{10.1016/j.trc.2020.102640}.
\newblock URL \url{https://www.sciencedirect.com/science/article/pii/S0968090X19306436}.

\bibitem[OpenStreetMap(2023)]{openstreetmap_taghighwaycycleway_2023}
OpenStreetMap.
\newblock Tag:highway=cycleway - {OpenStreetMap} {Wiki}, 2023.
\newblock URL \url{https://wiki.openstreetmap.org/wiki/Tag:highway%3Dcycleway}.

\bibitem[{OpenStreetMap Contributors}(2023)]{openstreetmap_contributors_openstreetmap_2023}
{OpenStreetMap Contributors}.
\newblock {OpenStreetMap}, 2023.
\newblock URL \url{https://www.openstreetmap.org/}.

\bibitem[OSM(2022)]{osm_openstreetmap_2022}
OSM.
\newblock {OpenStreetMap} for {Government}, 2022.
\newblock URL \url{https://wiki.openstreetmap.org/wiki/OpenStreetMap_for_Government}.

\bibitem[Paulsen and Rich(2023)]{paulsen_societally_2023}
Mads Paulsen and Jeppe Rich.
\newblock Societally optimal expansion of bicycle networks.
\newblock \emph{Transportation Research Part B: Methodological}, 174:\penalty0 102778, August 2023.
\newblock ISSN 0191-2615.
\newblock \doi{10.1016/j.trb.2023.06.002}.
\newblock URL \url{https://www.sciencedirect.com/science/article/pii/S0191261523000954}.

\bibitem[PeopleForBikes(2023)]{peopleforbikes_bna_2023}
PeopleForBikes.
\newblock {BNA} {Bicycle} {Network} {Analysis}, 2023.
\newblock URL \url{https://bna.peopleforbikes.org/#/}.

\bibitem[{Rambøll}(2022)]{ramboll_walking_2022}
{Rambøll}.
\newblock Walking and cycling data. {Practice}, challenges, needs and gaps, 2022.
\newblock URL \url{https://ramboll.com/-/media/files/rgr/documents/markets/transport/walking-cycling-data-gaps-2022.pdf}.

\bibitem[Reggiani et~al.(2021)Reggiani, van Oijen, Hamedmoghadam, Daamen, Vu, and Hoogendoorn]{reggiani_understanding_2021}
Giulia Reggiani, Tim van Oijen, Homayoun Hamedmoghadam, Winnie Daamen, Hai~L. Vu, and Serge Hoogendoorn.
\newblock Understanding bikeability: a methodology to assess urban networks.
\newblock \emph{Transportation}, June 2021.
\newblock ISSN 1572-9435.
\newblock \doi{10.1007/s11116-021-10198-0}.
\newblock URL \url{https://doi.org/10.1007/s11116-021-10198-0}.

\bibitem[Reggiani et~al.(2023)Reggiani, Verma, Daamen, and Hoogendoorn]{reggiani_multi-city_2023}
Giulia Reggiani, Trivik Verma, Winnie Daamen, and Serge Hoogendoorn.
\newblock A multi-city study on structural characteristics of bicycle networks.
\newblock \emph{Environment and Planning B: Urban Analytics and City Science}, page 23998083231170637, April 2023.
\newblock ISSN 2399-8083.
\newblock \doi{10.1177/23998083231170637}.
\newblock URL \url{https://doi.org/10.1177/23998083231170637}.
\newblock Publisher: SAGE Publications Ltd STM.

\bibitem[Rey and Anselin(2007)]{rey_pysal_2007}
Sergio Rey and Luc Anselin.
\newblock {PySAL}: {A} {Python} {Library} of {Spatial} {Analytical} {Methods}.
\newblock \emph{Review of Regional Studies}, January 2007.
\newblock \doi{10.52324/001c.8285}.

\bibitem[Rey et~al.(2020)Rey, Arribas-Bel, and Wolf]{rey_geographic_2020}
Sergio~J. Rey, Dani Arribas-Bel, and Levi~J. Wolf.
\newblock Geographic {Thinking} for {Data} {Scientists} — {Geographic} {Data} {Science} with {Python}, 2020.
\newblock URL \url{https://geographicdata.science/book/notebooks/01_geo_thinking.html}.

\bibitem[Saxton(2022)]{saxton_mapping_2022}
Tyler Saxton.
\newblock Mapping suburban bicycle lanes using street scene images and deep learning, April 2022.
\newblock URL \url{http://arxiv.org/abs/2204.12701}.
\newblock arXiv:2204.12701 [cs].

\bibitem[Schiavina et~al.(2023)Schiavina, Freire, and MacManus]{schiavina_ghs-pop_2023}
Marcello Schiavina, Sergio Freire, and Kytt MacManus.
\newblock {GHS}-{POP} {R2023A} - {GHS} population grid multitemporal (1975-2030), April 2023.
\newblock URL \url{http://data.europa.eu/89h/2ff68a52-5b5b-4a22-8f40-c41da8332cfe}.

\bibitem[Schoner and Levinson(2014)]{schoner_missing_2014}
Jessica~E. Schoner and David~M. Levinson.
\newblock The missing link: bicycle infrastructure networks and ridership in 74 {US} cities.
\newblock \emph{Transportation}, 41\penalty0 (6):\penalty0 1187--1204, November 2014.
\newblock ISSN 0049-4488, 1572-9435.
\newblock \doi{10.1007/s11116-014-9538-1}.
\newblock URL \url{http://link.springer.com/10.1007/s11116-014-9538-1}.

\bibitem[Septima(2019)]{septima_geodanmark_2019}
Septima.
\newblock {GeoDanmark} og ruteplanlægning.
\newblock Technical report, GeoDanmark, Copenhagen, 2019.

\bibitem[{Statistics Denmark}(2023)]{statistics_denmark_statistikbanken_2023}
{Statistics Denmark}.
\newblock Statistikbanken, 2023.
\newblock URL \url{https://statistikbanken.dk/folk1a}.

\bibitem[Steinacker et~al.(2022)Steinacker, Storch, Timme, and Schröder]{steinacker_demand-driven_2022}
Christoph Steinacker, David-Maximilian Storch, Marc Timme, and Malte Schröder.
\newblock Demand-driven design of bicycle infrastructure networks for improved urban bikeability.
\newblock \emph{Nature Computational Science}, pages 1--10, October 2022.
\newblock ISSN 2662-8457.
\newblock \doi{10.1038/s43588-022-00318-w}.
\newblock URL \url{https://www.nature.com/articles/s43588-022-00318-w}.
\newblock Publisher: Nature Publishing Group.

\bibitem[Szell et~al.(2022)Szell, Mimar, Perlman, Ghoshal, and Sinatra]{szell_growing_2022}
Michael Szell, Sayat Mimar, Tyler Perlman, Gourab Ghoshal, and Roberta Sinatra.
\newblock Growing urban bicycle networks.
\newblock \emph{Scientific Reports}, 12\penalty0 (1):\penalty0 6765, April 2022.
\newblock ISSN 2045-2322.
\newblock \doi{10.1038/s41598-022-10783-y}.
\newblock URL \url{https://www.nature.com/articles/s41598-022-10783-y}.
\newblock Number: 1 Publisher: Nature Publishing Group.

\bibitem[Tenkanen(2021)]{tenkanen_htenkanenpyrosm_2021}
Henrikki Tenkanen.
\newblock {HTenkanen}/pyrosm: v0.6.1, October 2021.
\newblock URL \url{https://zenodo.org/record/5561232}.

\bibitem[Uber(2023)]{uber_h3-py_2023}
Uber.
\newblock h3-py: {Uber}'s {H3} {Hexagonal} {Hierarchical} {Geospatial} {Indexing} {System} in {Python}, June 2023.
\newblock URL \url{https://github.com/uber/h3-py}.
\newblock original-date: 2018-06-12T22:39:59Z.

\bibitem[Vierø et~al.(2023)Vierø, Vybornova, and Szell]{viero_bikedna_2023}
Ane~Rahbek Vierø, Anastassia Vybornova, and Michael Szell.
\newblock {BikeDNA}: {A} tool for bicycle infrastructure data and network assessment.
\newblock \emph{Environment and Planning B: Urban Analytics and City Science}, page 23998083231184471, June 2023.
\newblock ISSN 2399-8083.
\newblock \doi{10.1177/23998083231184471}.
\newblock URL \url{https://doi.org/10.1177/23998083231184471}.
\newblock Publisher: SAGE Publications Ltd STM.

\bibitem[Vybornova et~al.(2022)Vybornova, Cunha, Gühnemann, and Szell]{vybornova_automated_2022}
Anastassia Vybornova, Tiago Cunha, Astrid Gühnemann, and Michael Szell.
\newblock Automated {Detection} of {Missing} {Links} in {Bicycle} {Networks}.
\newblock \emph{Geographical Analysis}, n/a\penalty0 (n/a), 2022.
\newblock ISSN 1538-4632.
\newblock \doi{10.1111/gean.12324}.
\newblock URL \url{https://onlinelibrary.wiley.com/doi/abs/10.1111/gean.12324}.

\bibitem[Wasserman et~al.(2019)Wasserman, Rixey, Zhou, Levitt, and Benjamin]{wasserman_evaluating_2019}
David Wasserman, Alex Rixey, Xinyi~(Elynor) Zhou, Drew Levitt, and Matt Benjamin.
\newblock Evaluating {OpenStreetMap}’s {Performance} {Potential} for {Level} of {Traffic} {Stress} {Analysis}.
\newblock \emph{Transportation Research Record}, 2673\penalty0 (4):\penalty0 284--294, April 2019.
\newblock ISSN 0361-1981.
\newblock \doi{10.1177/0361198119836772}.
\newblock URL \url{https://doi.org/10.1177/0361198119836772}.
\newblock Publisher: SAGE Publications Inc.

\bibitem[Will(2014)]{will_development_2014}
J.~Will.
\newblock \emph{Development of an automated matching algorithm to assess the quality of the {OpenStreetMap} road network: a case study in {Göteborg}, {Sweden}}.
\newblock PhD thesis, Lund University, 2014.
\newblock URL \url{https://www.semanticscholar.org/paper/Development-of-an-automated-matching-algorithm-to-%3A-Will/b3b77d579077b967820630db56522bef31654f21}.

\bibitem[Willberg et~al.(2021)Willberg, Tenkanen, Poom, Salonen, and Toivonen]{willberg_comparing_2021}
Elias Willberg, Henrikki Tenkanen, Age Poom, Maria Salonen, and Tuuli Toivonen.
\newblock Comparing spatial data sources for cycling studies: a review.
\newblock In \emph{Transport in {Human} {Scale} {Cities}}, pages 169--187. Edward Elgar Publishing, August 2021.
\newblock ISBN 978-1-80037-051-7.
\newblock URL \url{https://www.elgaronline.com/display/edcoll/9781800370500/9781800370500.00025.xml}.
\newblock Section: Transport in Human Scale Cities.

\bibitem[Winters et~al.(2022)Winters, Zanotto, and Butler]{winters_canadian_2022}
Meghan Winters, Moreno Zanotto, and Gregory Butler.
\newblock The {Canadian} {Bikeway} {Comfort} and {Safety} metrics ({Can}-{BICS}): {National} measures of the bicycling environment for use in research and policy.
\newblock \emph{Health Reports}, 33\penalty0 (82):\penalty0 13, 2022.

\bibitem[Wu and Kemp(2019)]{wu_global_2019}
An-Min Wu and Karen Kemp.
\newblock Global {Measures} of {Spatial} {Association}.
\newblock \emph{Geographic Information Science \& Technology Body of Knowledge}, 2019\penalty0 (Q1), January 2019.
\newblock ISSN 25772848.
\newblock \doi{10.22224/gistbok/2019.1.12}.
\newblock URL \url{https://gistbok.ucgis.org/bok-topics/global-measures-spatial-association}.

\bibitem[Xiao et~al.(2022)Xiao, Sluijs, Ogilvie, Patterson, and Panter]{xiao_shifting_2022}
Christina Xiao, Esther~van Sluijs, David Ogilvie, Richard Patterson, and Jenna Panter.
\newblock Shifting towards healthier transport: carrots or sticks? {Systematic} review and meta-analysis of population-level interventions.
\newblock \emph{The Lancet Planetary Health}, 6\penalty0 (11):\penalty0 e858--e869, November 2022.
\newblock ISSN 2542-5196.
\newblock \doi{10.1016/S2542-5196(22)00220-0}.
\newblock URL \url{https://www.sciencedirect.com/science/article/pii/S2542519622002200}.

\bibitem[Zhang and Ai(2015)]{zhang_how_2015}
Xiang Zhang and Tinghua Ai.
\newblock How to {Model} {Roads} in {OpenStreetMap}? {A} {Method} for {Evaluating} the {Fitness}-for-{Use} of the {Network} for {Navigation}.
\newblock In Francis Harvey and Yee Leung, editors, \emph{Advances in {Spatial} {Data} {Handling} and {Analysis}: {Select} {Papers} from the 16th {IGU} {Spatial} {Data} {Handling} {Symposium}}, Advances in {Geographic} {Information} {Science}, pages 143--162. Springer International Publishing, Cham, 2015.
\newblock ISBN 978-3-319-19950-4.
\newblock \doi{10.1007/978-3-319-19950-4_9}.
\newblock URL \url{https://doi.org/10.1007/978-3-319-19950-4_9}.

\bibitem[Zhang et~al.(2021)Zhang, Wang, Jiao, Zhou, Yu, and Cheng]{zhang_detecting_2021}
Xiang Zhang, Tianfu Wang, Delin Jiao, Zhiying Zhou, Jianwei Yu, and Xiao Cheng.
\newblock Detecting inconsistent information in crowd-sourced street networks based on parallel carriageways identification and the rule of symmetry.
\newblock \emph{ISPRS Journal of Photogrammetry and Remote Sensing}, 175:\penalty0 386--402, May 2021.
\newblock ISSN 0924-2716.
\newblock \doi{10.1016/j.isprsjprs.2021.03.014}.
\newblock URL \url{https://www.sciencedirect.com/science/article/pii/S092427162100085X}.

\end{thebibliography}

\end{document}